\documentclass[11pt]{article}
\usepackage{latexsym}
\usepackage{amsfonts}
\usepackage{amsmath}
\usepackage{amsthm} 
\usepackage{amssymb}
\usepackage{epsfig}

\def\nuc{\nu(\mathbf{c})}
\def\nucS{\nu(\mathbf{c},S)}
\def\TUMin{{\mathrm{TUmin}}({\mathbf c})}

\def\TUMax{{\mathrm{TUmax}}({\mathbf c})}

\def\NTUMin{{\mathrm{NTUmin}}({\mathbf c})}

\def\NTUMax{{\mathrm{NTUmax}}({\mathbf c})}

\def\TUMinS{{\mathrm{TUmin}}({\mathbf c},S)}
\def\TUMinPrimeS{{\mathrm{TUmin}}({\mathbf c}',S)}
\def\TUMaxS{{\mathrm{TUmax}}({\mathbf c},S)}
\def\TUMaxPrimeS{{\mathrm{TUmax}}({\mathbf c}',S)}
\def\NTUMinS{{\mathrm{NTUmin}}({\mathbf c},S)}
\def\NTUMinPrimeS{{\mathrm{NTUmin}}({\mathbf c}',S)}
\def\NTUMaxS{{\mathrm{NTUmax}}({\mathbf c},S)}
\def\NTUMaxPrimeS{{\mathrm{NTUmax}}({\mathbf c}',S)}
\def\ratioTUMin{\phi_{\mathrm{TUmin}}}
\def\ratioTUMax{\phi_{\mathrm{TUmax}}}
\def\ratioNTUMin{\phi_{\mathrm{NTUmin}}}
\def\ratioNTUMax{\phi_{\mathrm{NTUmax}}}
\def\VCG{\mathrm{VCG}}
\def\x{{\mathbf x}}
\def\y{{\mathbf y}}

\newtheorem{theorem}{Theorem}
\newtheorem{lemma}[theorem]{Lemma}
\newtheorem{remark}{Remark}
\newtheorem{corollary}[theorem]{Corollary}

\newtheorem{proposition}[theorem]{Proposition}
\newtheorem{claim}[theorem]{Claim}
\newtheorem{example}[theorem]{Example}

\title{Frugality Ratios And Improved Truthful Mechanisms for
      Vertex Cover\thanks{
This research is supported by the
EPSRC research grants ``Algorithmics of Network-sharing Games'' and
``Discontinuous Behaviour in the Complexity of randomized Algorithms''.}}

\author{Edith Elkind\\
School of Electronics and Computer Science,\\
University of Southampton,\\
Southampton SO17 1BJ, U.\,K.
\and
Leslie Ann Goldberg\\
Dept.\ of Computer Science\\
University of Liverpool,\\
Ashton Street, Liverpool L69 3BX, U.\,K.\\
\and
Paul W.\ Goldberg\\
Dept.\ of Computer Science\\
University of Liverpool,\\
Ashton Street, Liverpool L69 3BX, U.\,K.\\
}

\begin{document}

\maketitle

\thispagestyle{empty}

\begin{abstract}
In {\em set-system auctions}, there are several overlapping teams of
agents, and a task that can be completed by any of these teams.  The
buyer's goal is to hire a team and pay as little as possible.
Recently, Karlin, Kempe and Tamir introduced a new definition of {\em
frugality ratio} for this setting. Informally, the frugality ratio is
the ratio of the total payment of a mechanism to perceived fair
cost. In this paper, we study this together with alternative notions
of fair cost, and how the resulting frugality ratios relate to each
other for various kinds of set systems.

We propose a new truthful polynomial-time auction for the vertex cover
problem (where the feasible sets correspond to the vertex covers of a
given graph), based on the {\em local ratio} algorithm of Bar-Yehuda
and Even. The mechanism guarantees to find a winning set whose cost is
at most twice the optimal. In this situation, even though it is NP-hard to
find a lowest-cost feasible set, we show that {\em local
optimality} of a solution can be used to derive frugality bounds that
are within a constant factor of best possible.
To prove this result, we use our alternative notions of frugality
via a bootstrapping technique, which may be of independent interest.

\end{abstract}

\clearpage

\setcounter{page}{1}

\setcounter{secnumdepth}{5}

\section{Introduction}

The situations where one has to hire a team of agents to perform a
task are quite typical in many domains. In a market-based environment,
this goal can be achieved by means of a (combinatorial) procurement
auction: the agents submit their bids and the buyer selects a team
based on the agents' ability to work with each other as well as their
payment requirements. The problem is complicated by the fact that only
{\em some} subsets of agents constitute a valid team: the task may
require several skills, and each agent may possess only a subset of
these skills, the agents must be able to communicate with each other,
etc.  Also, for each agent there is a cost associated with performing
the task.  This cost is known to the agent himself, but not to the
buyer or other agents.

A well-known example of this setting is a {\em shortest path auction},
where the buyer wants to purchase connectivity between two points in a
network that consists of independent subnetworks. In this case, the
valid teams are sets of links that contain a path between these two
points.  This problem has been studied extensively in the recent
literature starting with the seminal paper by Nisan and
Ronen~\cite{nr} (see also~\cite{at,feig,ess,cr,immorlica,e1,rt}).
Generally, problems in this category can be formalized by specifying
(explicitly or implicitly) the sets of agents capable of performing
the tasks, or {\em feasible} sets. Consequently, the auctions of this
type are sometimes referred to as {\em set system auctions}.

The buyer and the agents have conflicting goals: the buyer wants to
spend as little money as possible, and the agents want to maximise
their earnings.  Therefore, to ensure truthful bidding, the buyer has
to use a carefully designed payment scheme. While it is possible to
use the celebrated VCG mechanism~\cite{v,c,g} for this purpose, it
suffers from two drawbacks. First, to use VCG, the buyer always has
to choose a cheapest feasible set.  If the problem of finding a
cheapest feasible set is computationally hard, this may require
exponential computational effort. One may hope to use approximation
algorithms to mitigate this problem: the buyer may be satisfied with
a feasible set whose cost is {\em close} to optimal and for many
NP-hard problems there exist fast algorithms for finding approximately
optimal solutions.  However, generally speaking, one cannot combine
such algorithms with VCG-style payments and preserve
truthfulness~\cite{nr2}. The second issue with VCG is that it has to
pay a bonus to each agent in the winning team. As a result, the total
VCG payment may greatly exceed the true cost of a cheapest feasible
set. In fact, one can easily construct an example where this is indeed
the case. While the true cost of a cheapest feasible set is not
necessarily a realistic benchmark for a truthful mechanism, it turns
out that VCG performs quite badly with respect to more natural
benchmarks discussed later in the paper.  Therefore, a natural
question to ask is whether one can design truthful mechnisms and
reasonable benchmarks for a given set system such that these
mechanisms perform well with respect to these benchmarks.

This issue was first raised by Nisan and Ronen~\cite{nr}. It was
subsequently addressed by Archer and Tardos~\cite{at}, who introduced
the concept of {\em frugality} in the context of shortest path
auctions. The paper~\cite{at} proposes to measure the overpayment of a
mechanism by the worst-case ratio between its total payment and the
cost of the cheapest path that is disjoint from the path selected by
the mechanism; this quantity is called the {\em frugality ratio}. The
authors show that for a large class of truthful mechanisms for this
problem (which includes VCG and all mechanisms that satisfy certain
natural properties) the frugality ratio is $n$, where $n$ is the
number of edges in the shortest path.  Subsequently, Elkind et
al.~\cite{ess} showed that a somewhat weaker bound of $n/2$ holds for
{\em all} truthful shortest path auctions.  Talwar~\cite{t} extends
the definition of frugality ratio given in~\cite{at} to general set
systems, and studies the frugality ratio of the VCG mechanism for many
specific set systems, such as minimum spanning trees or set covers.

\begin{figure}
\begin{center}
\epsfig{file=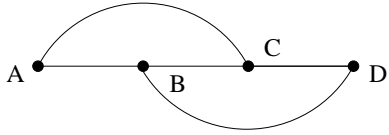, width=2in}
\end{center}
\caption{ The diamond graph }
\label{fig:diamond}
\end{figure}

While the definition of frugality ratio proposed by~\cite{at} is
well-motivated and has been instrumental in studying truthful
mechanisms for set systems, it is not completely
satisfactory. Consider, for example, the graph of
Figure~\ref{fig:diamond} with the costs $c_{AB}=c_{BC}=c_{CD}=0$,
$c_{AC}=c_{BD}=1$. This graph is 2-connected and the VCG payment to
the winning path ABCD is bounded. However, the graph contains no A--D
path that is disjoint from ABCD, and hence the frugality ratio of VCG
on this graph remains undefined. At the same time, there is no {\em
monopoly}, that is, there is no vendor that appears in all feasible
sets.  In auctions for other types of set systems, the requirement
that there exist a feasible solution disjoint from the selected one is
even more severe: for example, for vertex-cover auctions (where
vendors correspond to the vertices of some underlying graph, and the
feasible sets are vertex covers) the requirement means that the graph
must be bipartite.  To deal with this problem, Karlin et
al.~\cite{kkt} suggest a better benchmark, which is defined for any
monopoly-free set system. This quantity, which they denote by $\nu$,
intuitively corresponds to the total payoff in a cheapest Nash
equilibrium of a first-price auction.  Based on this new definition,
the authors construct new mechanisms for the shortest path problem and
show that the overpayment of these mechanisms is within a constant
factor of optimal.

\subsection{Our results}

\noindent{\bf Vertex cover auctions\ } We propose a truthful
polynomial-time auction for vertex cover that outputs a solution whose
cost is within a factor of 2 of optimal, and whose frugality ratio is
at most $2\Delta$, where $\Delta$ is the maximum degree of the graph
(Theorem~\ref{thm:2delta}).  We complement this result by proving
(Theorem~\ref{thm:delta/4}) that for any $\Delta$, there are
graphs of maximum degree $\Delta$ for which {\em
any} truthful mechanism has frugality ratio at least $\Delta/2$. This
means that both the solution quality and the frugality ratio of our
auction are within a constant factor of optimal.  In particular, the
frugality ratio is within a factor of~$4$ of optimal.  To the best of
our knowledge, this is the first auction for this problem that enjoys
these properties. Moreover, we show how to transform any truthful
mechanism for the vertex-cover problem into a frugal one while
preserving the approximation ratio.

\noindent{\bf Frugality ratios\ } Our vertex cover results naturally
suggest two modifications of the definition of $\nu$
in~\cite{kkt}. These modifications can be made independently of each
other, resulting in four different payment bounds that we denote as
$\mathrm{TUmax}$, $\mathrm{TUmin}$, $\mathrm{NTUmax}$, and
$\mathrm{NTUmin}$, where $\mathrm{NTUmin}$ is equal to the original
payment bound $\nu$ of in~\cite{kkt}.  All four payment bounds arise
as Nash equilibria of certain games (see Appendix); the differences
between them can be seen as ``the price of initiative'' and ``the
price of co-operation'' (see Section~\ref{sec:frugality}).  While our
main result about vertex cover auctions (Theorem~\ref{thm:2delta}) is
with respect to $\mathrm{NTUmin} =\nu$, we make use of the new
definitions by first comparing the payment of our mechanism to a
weaker bound $\mathrm{NTUmax}$, and then bootstrapping from this
result to obtain the desired bound.

Inspired by this application, we embark on a further study
of these payment bounds. Our results here are as follows:
\begin{enumerate}
\item
We observe (Proposition \ref{inequalities}) that the payment bounds we
consider always obey a particular order that is independent of the
choice of the set system and the cost vector, namely, $\mathrm{TUmin}
\le \mathrm{NTUmin} \le \mathrm{NTUmax} \le \mathrm{TUmax}$.  We
provide examples (Proposition~\ref{exvcone} and
Corollaries~\ref{exvctwo} and~\ref{exvcthree}) showing that for the
vertex cover problem any two consecutive bounds can differ by a factor
of $n-2$, where $n$ is the number of agents. We then show
(Theorem~\ref{thm:upper}) that this separation is almost optimal for
general set systems by proving that for any set system
$\mathrm{TUmax}/\mathrm{TUmin}\le n$.  In contrast, we demonstrate
(Theorem~\ref{thm:upperpath}) that for path auctions
$\mathrm{TUmax}/\mathrm{TUmin} \le 2$. We provide examples
(Proposition~\ref{expath})
showing that this bound is tight. We see this as an argument for the
study of vertex-cover auctions, as they appear to be more
representative of the general team-selection problem than the widely
studied path auctions.
\item
We show (Theorem~\ref{thm:ratios}) that for any set system, if there
is a cost vector for which $\mathrm{TUmin}$ and $\mathrm{NTUmin}$
differ by a factor of $\alpha$, there is another cost vector that
separates $\mathrm{NTUmin}$ and $\mathrm{NTUmax}$ by the same factor
and vice versa; the same is true for the pairs $(\mathrm{NTUmin},
\mathrm{NTUmax})$ and $(\mathrm{NTUmax}, \mathrm{TUmax})$. This
result suggests that the four payment bounds should be studied in a
unified framework; moreover, it leads us to believe that the
bootstrapping technique of Theorem~\ref{thm:2delta} may have other
applications.
\item
We evaluate the payment bounds introduced here with respect to a
checklist of desirable features.  In particular, we note that the
payment bound $\nu=\mathrm{NTUmin}$ of~\cite{kkt} exhibits some
counterintuitive properties, such as nonmonotonicity with respect to
adding a new feasible set (Proposition~\ref{clm:nm}), and is NP-hard
to compute (Theorem~\ref{thm:NPhard}), while some of the other payment
bounds do not suffer from these problems. This can be seen as an
argument in favour of using weaker but efficiently computable bounds
$\mathrm{NTUmax}$ and $\mathrm{TUmax}$.
\end{enumerate}

\subsection{Related work on vertex-cover auctions}

Vertex-cover auctions have been studied in the past by Talwar~\cite{t}
and Calinescu~\cite{cal}. Both of these papers are based on the
definition of frugality ratio used in~\cite{at}; as mentioned before,
this means that their results only apply to bipartite
graphs. Talwar~\cite{t} shows that the frugality ratio of VCG is at
most $\Delta$.  However, since finding the cheapest vertex cover is an
NP-hard problem, the VCG mechanism is computationally infeasible.  The
first (and, to the best of our knowledge, only) paper to investigate
polynomial-time truthful mechanisms for vertex cover
is~\cite{cal}. That paper studies an auction that is based on the
greedy allocation algorithm, which has an approximation ratio of $\log
n$. While the main focus of~\cite{cal} is the more general set cover
problem, the results of~\cite{cal} imply a frugality ratio of
$2\Delta^2$ for vertex cover. Our results improve on those of~\cite{t}
as our mechanism is polynomial-time computable, as well as on those
of~\cite{cal}, as our mechanism has a better approximation ratio, and
we prove a stronger bound on the frugality ratio; moreover, this bound
also applies to the mechanism of~\cite{cal}.

\section{Preliminaries}\label{sec:preliminaries}

A {\em set system} is a pair $(\mathcal{E},\mathcal{F})$, where
$\mathcal{E}$ is the {\em ground set}, $|\mathcal{E}|=n$, and
$\mathcal{F}$ is a collection of {\em feasible sets}, which are
subsets of $\mathcal{E}$.
Two particular types of set systems are of
particular interest to us --- {\em shortest path} systems
and \emph{vertex cover} systems. In a shortest path system,
the ground set consists of all edges of a network, and
a set of edges is feasible if it contains a path
between two specified vertices $s$ and $t$. In a vertex cover system,
the elements of the ground set are the
vertices of a graph, and the feasible sets are vertex covers of this
graph. We will also present some results for {\em matroid} systems,
in which the ground set is the set of all elements of a matroid, and
the feasible sets are the bases of the matroid. For a formal definition
of a matroid, the reader is referred to~\cite{ox}. In this paper, we
use the following characterisation of a matroid.

\begin{proposition}\label{matroid}
A collection of feasible sets $\mathcal{F}$ is the set of bases of a matroid
if and only if for any $S_i, S_j\in\mathcal{F}$, there is a bijection $f$
between $S_i\setminus S_j$ and $S_j\setminus S_i$ such that
$S_i\setminus\{e\}\cup\{f(e)\}\in\mathcal{F}$ for any $e\in S_i\setminus S_j$.
\end{proposition}

In set system auctions, each element $e$ of the ground set is owned by
an independent agent and has an associated non-negative cost $c_e$.
The goal of the buyer is to select (purchase) a feasible set.  Each
element $e$ in the selected set incurs a cost of $c_e$. The elements
that are not selected incur no costs.

The auction proceeds as follows: all elements of the ground set make
their bids, then the buyer selects a feasible set based on the bids and
makes payments to the agents. Formally, an auction is defined by an
{\em allocation rule} $A:\mathbf{R}^n\mapsto\mathcal{F}$ and a {\em
payment rule} $P:\mathbf{R}^n\mapsto\mathbf{R}^n$.  The allocation
rule takes as input a vector of bids and decides which of the sets in
$\mathcal{F}$ should be selected.  The payment rule also takes as
input a vector of bids and decides how much to pay to each agent.  The
standard requirements are {\em individual rationality}, that the
payment to each agent should be at least as high as its incurred cost
(0 for agents not in the selected set and $c_e$ for an agent $e$ in
the selected set), and {\em incentive compatibility}, or {\em
truthfulness}, that each agent's dominant strategy is to bid its true
cost.

An allocation rule is {\em monotone} if an agent cannot increase its
chance of getting selected by raising its bid.  Formally, for any bid
vector $\mathbf{b}=(b_1,\ldots,b_n)$ and any $e\in\mathcal{E}$, if
$e\not\in A(\mathbf{b})$ then $e\not\in A(b_1, \dots, b'_e, \dots,
b_n)$ for any $b'_e > b_e$.  Given a monotone allocation rule $A$ and
a bid vector $\mathbf{b}$, the {\em threshold bid} $t_e$ of an agent
$e\in A(\mathbf{b})$ is the highest bid of this agent that still wins
the auction, given that the bids of other participants remain the
same. Formally, $t_e=\sup\{b'_e\in\mathbf{R}\mid e\in A(b_1, \dots,
b'_e, \dots, b_n)\}$.  It is well known (see, e.g.~\cite{nr,ghw}) that
any set-system auction
that has a monotone allocation rule and pays each agent
its threshold bid is truthful; conversely, any truthful set-system
auction has a monotone allocation rule.

The VCG mechanism is a truthful mechanism that maximises the ``social
welfare'' and pays 0 to the losing agents.  For set system auctions,
this simply means picking a cheapest feasible set, paying each agent
in the selected set its threshold bid, and paying 0 to all other
agents.  Note, however, that the VCG mechanism may be difficult to
implement, since finding a cheapest feasible set may be computationally
hard.

If $U\subseteq \mathcal{E}$ is a set of agents, $c(U)$ denotes
$\sum_{e\in U} c_e$. (Note that we identify an agent with its
associated member of the ground set.) Similarly, $b(U)$ denotes
$\sum_{e \in U} b_e$.

\section{Frugality ratios}
\label{sec:frugality}

We start by reproducing the definition of the quantity $\nu$
from~\cite[Definition 4]{kkt}.  Let $(\mathcal{E},\mathcal{F})$ be a
set system and let $S$ be a cheapest feasible set with respect to the
(vector of) true costs $\mathbf{c}$. Then $\nucS$ is the solution to
the following optimisation problem.

Minimise $B = \sum_{e\in S} b_e$ subject to
\begin{itemize}
\item[(1)] $b_e \ge c_e$ for all $e\in S$
\item[(2)] $\sum_{e\in S\setminus T} b_e \leq \sum_{e \in T\setminus
S}c_e$ for all $T\in \mathcal{F}$
\item[(3)] for every $e\in S$, there is $T_e\in \mathcal{F}$ such
that $e\not\in T_e$ and $\sum_{e'\in S\setminus T_e} b_{e'} = \sum_{e'
\in T_e \setminus S} c_{e'}$
\end{itemize}

The bound~$\nucS$ can be seen as an outcome of a hypothetical
two-stage process as follows. An omniscient auctioneer knows all
the vendors' private costs, and identifies a cheapest set $S$.
The auctioneer offers payments to the members of $S$.
He does it so as to minimise his total payment subject to
the following constraints that represent a notion of fairness.
\begin{itemize}
\item The payment to any member of $S$ covers that member's cost.
(Condition 1)
\item $S$ is still a cheapest set with respect to the new cost vector in
which the cost of a member of $S$ has been increased to his offer.
(Condition 2)
\item if any member $e$ of $S$ were to ask for a higher payment than
his offer, then some other feasible set (not containing $e$) would be
cheapest.  (Condition 3)
\end{itemize}

This definition captures many important aspects of our intuition about
`fair' payments. However, it can be modified in two ways, both of
which are still quite natural, but result in different payment bounds.

First, we can consider the worst rather than the best possible outcome
for the buyer. That is, we can consider the maximum total payment that
the agents can extract by jointly selecting their bids subject to (1),
(2), and (3).  Such a bound corresponds to maximising $B$ subject to
(1), (2), and (3) rather than minimising it. If the agents in $S$
submit bids (rather than the auctioneer making offers), this kind of
outcome is plausible. It has to be assumed that agents submit bids
independently of each other, but know how high they can bid and still
win.  Hence, the difference between these two definitions can be seen
as ``the price of initiative''.

Second, the agents may be able to make payments to each other.  In
this case, if they can extract more money from the buyer by agreeing
on a vector of bids that violates individual rationality (i.e.,
condition~(1)) for some bidders, they might be willing to do so, as
the agents who are paid below their costs will be compensated by other
members of the group.  The bids must still be realistic, i.e., they
have to satisfy $b_e\ge 0$.  The resulting change in payments can be
seen as ``the price of co-operation'' and corresponds to replacing
condition~(1) with the following weaker condition~$(1^*)$:
\begin{equation*}
b_e\ge 0\text{ for all } e\in S.
\tag{$1^*$}
\end{equation*}

By considering all possible combinations of these modifications,
we obtain four different payment bounds, namely

\begin{itemize}
\item $\TUMinS$, which is the
solution to the optimisation problem ``Minimise $B$
subject to $(1^*)$, (2), and (3)''.
\item $\TUMaxS$, which is the solution to the optimisation problem
``Maximise $B$ subject to $(1^*)$, (2), and (3)''.
\item $\NTUMinS$, which is the solution to the optimisation problem
``Minimise $B$ subject to (1), (2), and (3)''.
\item $\NTUMaxS$, which is the solution to the optimisation problem
``Maximise $B$ subject to (1), (2), (3)''.
\end{itemize}
The abbreviations TU and NTU correspond, respectively, to transferable
utility and non-transferable utility, i.e., the agents'
ability/inability to make payments to each other.  For concreteness,
we will take $\TUMin$ to be $\TUMinS$ where $S$ is the
lexicographically least amongst the cheapest feasible sets.  We define
$\TUMax$, $\NTUMin$, $\NTUMax$ and $\nuc$ similarly, though we will
see in Section~\ref{sec:choiceS} that $\NTUMinS$ and $\NTUMaxS$ are
independent of the choice of~$S$.  Note that the quantity
$\nu({\mathbf c})$ from \cite{kkt} is $\NTUMin$.

The second modification (transferable utility) is more intuitively
appealing in the context of the maximisation problem, as both assume
some degree of co-operation between the agents.  While the second
modification can be made without the first, the resulting payment
bound $\TUMinS$ turns out to be too strong to be a realistic
benchmark, at least for general set systems. In particular, it can be
smaller than the total cost of a cheapest feasible set $S$ (see
Section~\ref{sec:properties}). However, we provide the definition and
some results about $\TUMinS$, both for completeness and because we
believe that it may help to understand which properties of the payment
bounds are important for our proofs.  Another possibility would be to
introduce an additional constraint $\sum_{e\in S}b_e\ge \sum_{e\in
S}c_e$ in the definition of $\TUMinS$ (note that this condition holds
automatically for non-transferable utility bounds,
and also for $\TUMaxS$, as $\TUMaxS\ge\NTUMaxS$). However, such a
definition would have no direct economic interpretation, and some of
our results (in particular, the ones in Section~\ref{sec:compare})
would no longer hold.

\begin{remark}\label{maxeasy}
For the payment bounds that are derived from maximisation problems,
(i.e., $\TUMaxS$ and $\NTUMaxS$), constraints of type (3) are
redundant and can be dropped. Hence, $\TUMaxS$ and $\NTUMaxS$ are
solutions to linear programs, and therefore can be computed in
polynomial time as long as we have a separation oracle for constraints
in~(2). In contrast, $\NTUMinS$ can be NP-hard to compute even if the
size of ${\cal F}$ is polynomial (see Section~\ref{sec:properties}).
\end{remark}

The first and third inequalities in the following observation follow
from the fact that condition $(1^*)$ is weaker than condition (1).

\begin{proposition}\label{inequalities}
$\TUMinS \leq \NTUMinS \leq \NTUMaxS\leq \TUMaxS$.
\end{proposition}

Let $\mathcal{M}$ be a truthful mechanism for
$(\mathcal{E},\mathcal{F})$.  Let $p_{\mathcal{M}}({\mathbf c})$
denote the total payments of $\mathcal{M}$ when the actual costs are
${\mathbf c}$.  A {\em frugality ratio} of $\mathcal{M}$ with respect
to a payment bound is the ratio between the payment of $\mathcal{M}$
and this payment bound. In particular,
\begin{align*}
\ratioTUMin(\mathcal{M})
 &= \sup_{\mathbf c} p_{\mathcal{M}}({\mathbf c})/\TUMin,\\
\ratioTUMax(\mathcal{M})
 &= \sup_{\mathbf c} p_{\mathcal{M}}({\mathbf c})/\TUMax,\\
\ratioNTUMin(\mathcal{M})
 &= \sup_{\mathbf c} p_{\mathcal{M}}({\mathbf c})/\NTUMin,\\
\ratioNTUMax(\mathcal{M})
 &= \sup_{\mathbf c} p_{\mathcal{M}}({\mathbf c})/\NTUMax.
\end{align*}
We conclude this section by showing that there exist set systems and
respective cost vectors for which all four payment bounds are
different. In the next section, we quantify this difference, both for
general set systems, and for specific types of set systems, such as
path auctions or vertex cover auctions.

\begin{example}
Consider the shortest-path auction on the graph of
Figure~\ref{fig:diamond}.  The minimal feasible sets are all paths from $A$ to
$D$.  It can be verified, using the reasoning of
Proposition~\ref{expath} below, that for the
cost vector $c_{AB}=c_{CD}=2$, $c_{BC}=1$, $c_{AC}=c_{BD}=5$, we have
\begin{itemize}
\item
$\TUMax=10$ (with the bid vector $b_{AB}=b_{CD}=5$, $b_{BC}=0$),
\item
$\NTUMax=9$ (with the bid vector $b_{AB}=b_{CD}=4$, $b_{BC}=1$),
\item
$\NTUMin=7$ (with the bid vector $b_{AB}=b_{CD}=2$, $b_{BC}=3$),
\item
$\TUMin=5$ (with the bid vector $b_{AB}=b_{CD}=0$, $b_{BC}=5$).
\end{itemize}
\end{example}

\section{Comparing payment bounds}\label{sec:compare}

\subsection{Path auctions}\label{sec:pathlb}

We start by showing that for path auctions any two consecutive payment
bounds (in the sequence of Proposition~\ref{inequalities}) can differ
by at least a factor of 2. In Section~\ref{sec:upper}
(Theorem~\ref{thm:upperpath}), we show that the separation results in
Proposition~\ref{expath} are optimal (that is, the factor of 2 is
maximal for path auctions).

\begin{proposition}\label{expath}
There is a path auction and cost vectors $\mathbf{c}$,
$\mathbf{c}'$ and $\mathbf{c}''$ for which
\begin{itemize}
\item[(i)] $\NTUMax/\NTUMin\ge 2$,
\item[(ii)] ${\rm TUmax}(\mathbf{c}')/{\rm NTUmax}(\mathbf{c}')\ge 2$,
\item[(iii)] ${\rm NTUmin}(\mathbf{c}'')/{\rm TUmin}(\mathbf{c}'')\ge 2$.
\end{itemize}
\end{proposition}

\begin{proof}
Consider the graph of Figure~\ref{fig:diamond}. For the cost vectors
$\mathbf{c}$, $\mathbf{c}'$ and $\mathbf{c}''$ defined below, ABCD
is the lexicographically-least cheapest path, so we can assume that
$S=\{AB, BC, CD\}$.
\begin{itemize}
\item[(i)]{
Let $\mathbf{c}$ be edge costs $c_{AB}=c_{BC}=c_{CD}=0$,
$c_{AC}=c_{BD}=1$.  The inequalities in~(2) are $b_{AB}+b_{BC}\le
c_{AC}=1$, $b_{BC}+b_{CD}\le c_{BD}=1$.  By condition~(3), both of
these inequalities must be tight (the former one is the only
inequality involving $b_{AB}$, and the latter one is the only
inequality involving $b_{CD}$).  The inequalities in~(1) are
$b_{AB}\ge 0$, $b_{BC}\ge 0$, $b_{CD}\ge 0$.  Now, if the goal is to
maximise $b_{AB}+b_{BC}+b_{CD}$, the best choice is $b_{AB}=b_{CD}=1$,
$b_{BC}=0$, so $\NTUMax=2$.  On the other hand, if the goal is to
minimise $b_{AB}+b_{BC}+b_{CD}$, one should set $b_{AB}=b_{CD}=0$,
$b_{BC}=1$, so $\NTUMin=1$.}

\item[(ii)]{
Let $\mathbf{c}'$ be the edge costs be $c'_{AB}=c'_{CD}=0$,
$c'_{BC}=1$, $c'_{AC}=c'_{BD}=1$.  The inequalities in~(2) are the
same as before, and by the same argument both of them
are, in fact, equalities.  The inequalities in~(1) are $b_{AB}\ge 0$,
$b_{BC}\ge 1$, $b_{CD}\ge 0$.  Our goal is to maximise
$b_{AB}+b_{BC}+b_{CD}$.  If we have to respect the inequalities
in~(1), we have to set $b_{AB}=b_{CD}=0$, $b_{BC}=1$, so
$\NTUMax=1$. Otherwise, we can set $b_{AB}=b_{CD}=1$, $b_{BC}=0$, so
$\TUMax\ge 2$.}

\item[(iii)]{
The edge costs $\mathbf{c}''$ are $c''_{AB}=c''_{CD}=1$, $c''_{BC}=0$,
$c''_{AC}=c''_{BD}=1$. Again, the inequalities in~(2) are the same,
and both are, in fact, equalities.  The inequalities in~(1) are
$b_{AB}\ge 1$, $b_{BC}\ge 0$, $b_{CD}\ge 1$.  Our goal is to minimise
$b_{AB}+b_{BC}+b_{CD}$.  If we have to respect the inequalities
in~(1), we have to set $b_{AB}=b_{CD}=1$, $b_{BC}=0$, so
$\NTUMin=2$. Otherwise, we can set $b_{AB}=b_{CD}=0$, $b_{BC}=1$, so
$\TUMin\le 1$.}
\end{itemize}
\end{proof}

\subsection{Connections between separation results}

The separation results for path auctions are obtained on the same
graph using very similar cost vectors. It turns out that this is not
coincidental. Namely, we can prove the following theorem.

\begin{theorem}\label{thm:ratios}
For any set system $(\mathcal{E}, \mathcal{F})$, and any feasible
set~$S$,
\[
\max_{\mathbf{c}}\frac{\TUMaxS}{\NTUMaxS}=
\max_{\mathbf{c}}\frac{\NTUMaxS}{\NTUMinS}=
\max_{\mathbf{c}}\frac{\NTUMinS}{\TUMinS},
\]
where the maximum is over all cost vectors~$\mathbf{c}$ for which $S$
is a cheapest feasible set.
\end{theorem}

The proof of the theorem follows directly from the four lemmas proved
below; in particular, the first equality in Theorem~\ref{thm:ratios}
is obtained by combining Lemmas~\ref{useone} and~\ref{usetwo}, and the
second equality is obtained by combining Lemmas~\ref{usethree}
and~\ref{usefour}.

\begin{lemma}\label{useone}
Suppose that $\mathbf{c}$ is a cost vector for $(\mathcal{E},
\mathcal{F})$ such that $S$ is a cheapest feasible set and\\
$\TUMaxS/\NTUMaxS=\alpha$. Then there is a cost vector $\mathbf{c}'$
such that $S$ is a cheapest feasible set and
$\NTUMaxPrimeS/\NTUMinPrimeS\ge\alpha$.
\end{lemma}

\begin{proof}
Suppose that $\TUMaxS=X$ and $\NTUMaxS=Y$ where $X/Y=\alpha$. Assume
without loss of generality that $S$ consists of elements $1, \dots,
k$, and let $\mathbf{b}^1=(b_1^1, \dots, b_k^1)$ and
$\mathbf{b}^2=(b_1^2, \dots, b_k^2)$ be the bid vectors that
correspond to $\TUMaxS$ and $\NTUMaxS$, respectively.

Construct the cost vector $\mathbf{c}'$ by setting $c'_i=c_i$ for
$i\not\in S$, $c'_i=\min\{c_i, b^1_i\}$ for $i\in S$.  Clearly, $S$ is
a cheapest set under $\mathbf{c}'$.  Moreover, as the costs of
elements outside of $S$ remain the same, the right-hand sides of all
constraints in~(2) and~(3) do not change, so any bid vector that
satisfies~(2) and~(3) with respect to $\mathbf{c}$, also satisfies
them with respect to $\mathbf{c}'$.  We will construct two bid vectors
$\mathbf{b}^3$ and $\mathbf{b}^4$ that satisfy conditions~(1), (2)
and~(3) for the cost vector $\mathbf{c}'$, and also satisfy
$\sum_{i\in S}b^3_i=X$, $\sum_{i\in S}b^4_i=Y$. It follows that
$\NTUMaxPrimeS\ge X$ and $\NTUMinPrimeS\le Y$, which implies the
lemma.

We can set $\mathbf{b}^3 = \mathbf{b}^1$: this bid vector satisfies
conditions~(2) and~(3) since $\mathbf{b}^1$ does, and we have
$b^1_i\ge \min\{c_i, b^1_i\}=c'_i$, which means that $\mathbf{b}^3$
satisfies condition~(1).  We can set $\mathbf{b}^4 =
\mathbf{b}^2$. Again, $\mathbf{b}^4$ satisfies conditions~(2) and~(3)
since $\mathbf{b}^2$ does, and since $\mathbf{b}^2$ satisfies
condition~(1), we have $b^2_i\ge c_i\ge c'_i$, which means that
$\mathbf{b}^4$ satisfies condition~(1).
\end{proof}

\begin{lemma}\label{usetwo}
Suppose that $\mathbf{c}$ is a cost vector for $(\mathcal{E},
\mathcal{F})$ such that $S$ is a cheapest feasible set and\\
$\NTUMaxS/\NTUMinS=\alpha$. Then there is a cost vector $\mathbf{c}'$
such that $S$ is a cheapest feasible set and
$\TUMaxPrimeS/\NTUMaxPrimeS\ge\alpha$.
\end{lemma}

\begin{proof}
Suppose that $\NTUMaxS=X$ and $\NTUMinS=Y$ where $X/Y=\alpha$. Again,
assume that $S$ consists of elements $1, \dots, k$, and let
$\mathbf{b}^1=(b_1^1, \dots, b_k^1)$ and $\mathbf{b}^2=(b_1^2, \dots,
b_k^2)$ be the bid vectors that correspond to $\NTUMaxS$ and
$\NTUMinS$, respectively.

Construct the cost vector $\mathbf{c}'$ by setting $c'_i=c_i$ for
$i\not\in S$, $c'_i=b^2_i$ for $i\in S$.  As $\mathbf{b}^2$ satisfies
condition~(2), $S$ is a cheapest set under $\mathbf{c}'$.  As in the
previous construction, the right-hand sides of all constraints in~(2)
do not change.  Let $\mathbf{b}^3$ be a bid vector that corresponds
to $\NTUMaxPrimeS$.  Let us prove that $\NTUMaxPrimeS=\sum_{i\in
S}b^3_i=Y$. Indeed, the bid vector $\mathbf{b}^3$ must satisfy
$b_i^3\ge c'_i=b^2_i$ for $i=1, \dots, k$ (condition~(1)).  Suppose
that $b_i^3 > c'_i$ for some $i=1, \dots, k$, and consider the
constraint in~(2) that is tight for $b^2_i$. There is such a
constraint, as $\mathbf{b}^2$ satisfies condition~(3). Namely, for
some $T$ not containing~$i$,
$$\sum_{j\in S\setminus T} b_j^2 = \sum_{j\in T\setminus S} c_j.$$
For every $j$ appearing in the left-side of this constraint,
we have $b^3_{j} \ge b^2_{j}$
but $b^3_i>b^2_i$, so
the bid vector $\mathbf{b}^3$ violates
this constraint. Hence, $b_i^3=c'_i=b^2_i$ for all $i$ and therefore
$\NTUMaxPrimeS=\sum_{i\in S}b^3_i= Y$.

On the other hand, we can construct a bid vector $\mathbf{b}^4$ that
satisfies conditions~(2) and~(3) with respect to $\mathbf{c}'$ and has
$\sum_{i\in S}b^4_i=X$. Namely, we can set $\mathbf{b}^4 =
\mathbf{b}^1$: as $\mathbf{b}^1$ satisfies conditions~(2) and~(3), so
does $\mathbf{b}^4$.  As $\TUMaxPrimeS\ge\sum_{i\in S}b^4_i$, this
proves the lemma.
\end{proof}

\begin{lemma}\label{usethree}
Suppose that $\mathbf{c}$ is a cost vector for $(\mathcal{E},
\mathcal{F})$ such that $S$ is a cheapest feasible set and\\
$\NTUMaxS/\NTUMinS=\alpha$. Then there is a cost vector $\mathbf{c}'$
such that $S$ is a cheapest feasible set and
$\NTUMinPrimeS/\TUMinPrimeS\ge\alpha$.
\end{lemma}

\begin{proof}
Suppose that $\NTUMaxS=X$ and $\NTUMinS=Y$ where $X/Y=\alpha$. Again,
assume $S$ consists of elements $1, \dots, k$, and let
$\mathbf{b}^1=(b_1^1, \dots, b_k^1)$ and $\mathbf{b}^2=(b_1^2, \dots,
b_k^2)$ be the bid vectors that correspond to $\NTUMaxS$ and
$\NTUMinS$, respectively.

The cost vector $\mathbf{c}'$ is obtained by setting $c'_i=c_i$ for
$i\not\in S$, $c'_i=b^1_i$ for $i\in S$.  Since $\mathbf{b}^1$
satisfies condition~(2), $S$ is a cheapest set under $\mathbf{c}'$,
and the right-hand sides of all constraints in~(2) do not change.

Let $\mathbf{b}^3$ be a bid vector that corresponds to
$\NTUMinPrimeS$.  It is easy to see that $\NTUMinPrimeS=\sum_{i\in
S}b^3_i=X$, since the bid vector $\mathbf{b}^3$ must satisfy $b_i^3\ge
c'_i=b^1_i$ for $i=1, \dots, k$ (condition~(1)), and $\sum_{i\in
S}b^1_i=\NTUMax=X$.  On the other hand, we can construct a bid vector
$\mathbf{b}^4$ that satisfies conditions~(2) and~(3) with respect to
$\mathbf{c}'$ and has $\sum_{i\in S}b^4_i=Y$. Namely, we can set
$\mathbf{b}^4 = \mathbf{b}^2$: as $\mathbf{b}^2$ satisfies
conditions~(2) and~(3), so does $\mathbf{b}^4$.  As
$\TUMinPrimeS\le\sum_{i\in S}b^4_i$, this proves the lemma.
\end{proof}

\begin{lemma}\label{usefour}
Suppose that $\mathbf{c}$ is a cost vector for $(\mathcal{E},
\mathcal{F})$ such that $S$ is a cheapest feasible set and\\
$\NTUMinS/\TUMinS=\alpha$. Then there is a cost vector $\mathbf{c}'$
such that $S$ is a cheapest feasible set and
$\NTUMaxPrimeS/\NTUMinPrimeS\ge\alpha$.
\end{lemma}

\begin{proof}
Suppose that $\NTUMinS=X$ and $\TUMinS=Y$ where $X/Y=\alpha$. Again,
assume that $S$ consists of elements $1, \dots, k$, and let
$\mathbf{b}^1=(b_1^1, \dots, b_k^1)$ and $\mathbf{b}^2=(b_1^2, \dots,
b_k^2)$ be the bid vectors that correspond to $\NTUMinS$ and
$\TUMinS$, respectively.

Construct the cost vector $\mathbf{c}'$ by setting $c'_i=c_i$ for
$i\not\in S$, $c'_i=\min\{c_i, b^2_i\}$ for $i\in S$.  Clearly, $S$ is
a cheapest set under $\mathbf{c}'$.  Moreover, as the costs of
elements outside of $S$ remained the same, the right-hand sides of all
constraints in~(2) do not change.

We construct two bid vectors $\mathbf{b}^3$ and $\mathbf{b}^4$
that satisfy conditions~(1), (2), and~(3) for the cost vector
$\mathbf{c}'$, and have $\sum_{i\in S}b^3_i=X$, $\sum_{i\in
S}b^4_i=Y$.  As $\NTUMaxPrimeS\ge X$ and $\NTUMinPrimeS\le Y$, this
implies the lemma.

We can set $\mathbf{b}^3=\mathbf{b}^1$. Indeed, the vector
$\mathbf{b}^3$ satisfies conditions~(2) and~(3) since $\mathbf{b}^1$
does. Also, since $\mathbf{b}^1$ satisfies condition~(1), we have
$b^1_i\ge c_i\ge c'_i$, i.e., $\mathbf{b}^3$ satisfies condition~(1)
with respect to $\mathbf{c}'$.  On the other hand, we can set
$\mathbf{b}^4 = \mathbf{b}^2$: the vector $\mathbf{b}^4$ satisfies
conditions~(2) and~(3) since $\mathbf{b}^2$ does, and it satisfies
condition~(1), since $b^2_i\ge c'_i$.
\end{proof}

\subsection{Vertex-cover auctions}
\label{sec:VC}

In contrast to the case of path auctions, for vertex-cover auctions
the gap between $\NTUMin$ and $\NTUMax$ (and hence between $\NTUMax$
and $\TUMax$, and between $\TUMin$ and $\NTUMin$) can be proportional
to the size of the graph.

\begin{figure}
\begin{center}
\epsfig{file=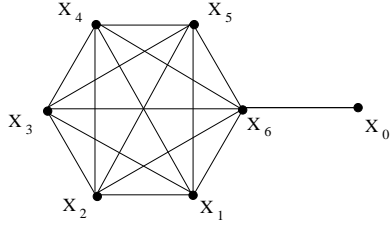, width=2in}
\end{center}
\caption{Graph that separates payment bounds for vertex cover, $n=7$}
\label{fig:vc}
\end{figure}

\begin{proposition}\label{exvcone}
For any $n\geq 3$, there is a an $n$-vertex graph and a cost vector
${\mathbf c}$ for which\\ $\TUMax/\NTUMax\ge n-2$.
\end{proposition}

\begin{proof}
The underlying graph consists of an $(n-1)$-clique on the vertices
$X_1, \dots, X_{n-1}$, and an extra vertex $X_0$ adjacent to
$X_{n-1}$. See Figure~\ref{fig:vc}.
The costs are $c_{X_1}=c_{X_2}=\dots=c_{X_{n-2}}=0$,
$c_{X_0}=c_{X_{n-1}}=1$.  We can assume that $S=\{X_0, X_1,\dots,
X_{n-2}\}$ (this is the lexicographically first vertex cover of
cost~$1$).  For this set system, the constraints in~(2) are
$b_{X_i}+b_{X_0}\leq c_{X_{n-1}}=1$ for $i=1, \dots, n-2$.  Clearly,
we can satisfy conditions~(2) and~(3) by setting $b_{X_i}=1$ for $i=1,
\dots, n-2$, $b_{X_0}=0$.  Hence, $\TUMax\ge n-2$.  For $\NTUMax$,
there is an additional constraint $b_{X_0}\ge 1$, so the best we can
do is to set $b_{X_i}=0$ for $i=1, \dots, n-2$, $b_{X_0}=1$, which
implies $\NTUMax=1$.
\end{proof}

Combining Proposition~\ref{exvcone} with Lemmas~\ref{useone}
and~\ref{usethree} (and re-naming vertices to make $S$ the
lexicographically-least cheapest feasible set), we derive the
following corollaries.

\begin{corollary}\label{exvctwo}
For any $n\geq 3$, there is an instance of the vertex cover problem on
an $n$-vertex graph for which for which $\NTUMax/\NTUMin\ge n-2$.
\end{corollary}

\begin{corollary}\label{exvcthree}
For any $n\geq 3$, there is an instance of the vertex cover problem on
an $n$-vertex graph for which $\NTUMin/\TUMin\ge n-2$.
\end{corollary}

\subsection{Upper bounds}\label{sec:upper}

It turns out that the lower bound proved in the previous subsection is
almost tight. More precisely, the following theorem shows that no two
payment bounds can differ by more than a factor of $n$; moreover, this
is the case not just for the vertex cover problem, but for general set
systems. We bound the gap between $\TUMax$ and $\TUMin$. Since
$\TUMin\le\NTUMin\le\NTUMax\le\TUMax$, this bound applies to any pair
of payment bounds.

\begin{theorem}\label{thm:upper}
For any set system auction having $n$ vendors and any cost vector
$\mathbf{c}$,
\[
\TUMax/\TUMin\le n.
\]
\end{theorem}

\begin{proof}
Let $k$ be the size of the winning set $S$. Let $c_1, \dots, c_k$ be
the true costs of elements in $S$, let $b_1, \dots, b_k$ be their bids
that correspond to $\TUMin$, and let $b'_1, \dots, b'_k$ be their bids
that correspond to $\TUMax$.  For $e'\in S$, let $T_{e'}$ be the
feasible set associated with $e'$ using (3) applied to the TUmin bids.

Since $\TUMin =\sum_{e\in S} b_e$, it follows that
\[
\begin{tabular}{rclr}
$k\TUMin$ & $=$    & $\sum_{e'\in S} \sum_{e\in S} b_e$ &  \\
          & $\geq$ & $\sum_{e'\in S} \sum_{e\in S\setminus T_{e'}} b_e$ & \\
          & $=$    & $\sum_{e'\in S} \sum_{e\in T_{e'}\setminus S} c_e$  &
~~~~~~~~by (3) (applied to TUmin bids) \\
          & $\geq$ & $\sum_{e'\in S} \sum_{e\in S\setminus T_{e'}} b'_e$ &
by (2) (applied to TUmax bids) \\
          & $\geq$ & $\sum_{e'\in S} b'_{e'}$  &
since $e' \in S\setminus T_{e'}$ \\
          & $=$    & $\TUMax$ & \\
\end{tabular}
\]
Since $k<n$ the result follows.
\end{proof}

\begin{remark}
The final line of the proof of Theorem~\ref{thm:upper} shows that, in
fact, the upper bound on\\ $\TUMax/\TUMin$ can be strengthened to the
size of the winning set, $k$.  Note that in Proposition~\ref{exvcone},
as well as in Corollaries~\ref{exvctwo} and~\ref{exvcthree}, $k=n-1$,
so these results do not contradict each other.
\end{remark}

For path auctions, this upper bound can be improved to 2, matching the
lower bounds of Section~\ref{sec:pathlb}.

\begin{figure}
\begin{center}
\epsfig{file=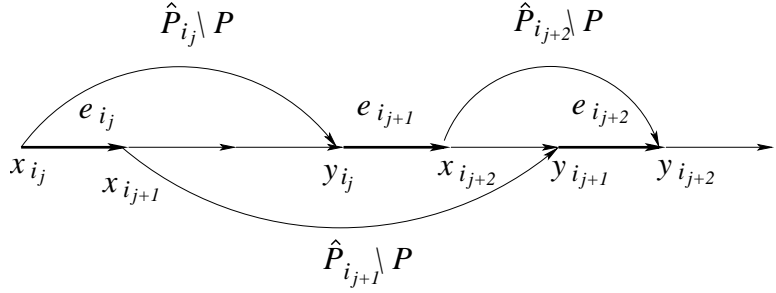, width=4in}
\end{center}
\caption{Proof of Theorem~\ref{thm:upperpath}:
constraints for $\hat{P}_{i_j}$ and $\hat{P}_{i_{j+2}}$ do not overlap}
\label{fig:avoid}
\end{figure}

\begin{theorem}\label{thm:upperpath}
For any path auction with cost vector $\mathbf{c}$,
${\TUMax} / \TUMin \le 2$.
\end{theorem}

\begin{proof}
Given a network $(G, s, t)$, let $P=\{e_1, \dots, e_k\}$ be
the le\-xi\-co\-gra\-phi\-cally-least cheapest $s$--$t$ path in $G$.
To simplify notation, relabel the vertices of $G$ as
$\{1, \dots, n\}$ so that
$e_1=(s, 1), e_2=(1, 2), \dots, e_k=(k-1, t)$.
Let $\mathbf{b}=(b_1,\dots,b_k)$ and
$\mathbf{b}'=(b'_1,\dots,b'_k)$ be bid vectors that correspond to
$\TUMin$ and $\TUMax$, respectively.

For $i=1,\ldots,k$ let $P_i$ be a $s$--$t$ path associated with
$e_i$ by a constraint of type~(3) applied to $\mathbf{b}$; consequently
$b(P\setminus P_i)=c(P_i\setminus P)$.
We can assume without loss of generality
that $P_i$ coincides with $P$ up to some vertex $x_i$, then deviates
from $P$ to avoid $e_i$, and finally returns to $P$ at a vertex $y_i$
and coincides with $P$ from then on (clearly, it might happen that
$s=x_i$ or $t=y_i$). Indeed, if $P_i$ deviates from $P$ more than
once, one of these deviations is not necessary to avoid $e_i$ and can
be replaced with the respective segment of $P$ without increasing the
cost of $P_i$. Among all paths of this form, let $\hat{P}_i$ be the
one with the largest value of $y_i$, i.e., the ``rightmost'' one. This
path corresponds to an equality~$I_i$ of the form
$b_{x_{i}+1}+\dots+b_{y_i} = c(\hat{P}_i\setminus P)$.

We construct a set $\mathcal{L}$ of equalities such that every
variable $b_i$ appears in at least one of them. We construct
$\mathcal{L}$ inductively as follows.  Start by setting
$\mathcal{L}=\{I_1\}$. At the $j$th step, suppose that all variables
up to (but not including) $b_{i_j}$ appear in at least one equality in
$\mathcal{L}$. Add $I_{i_j}$ to $\mathcal{L}$.

Note that for any $j$ we have $y_{i_{j+1}}>y_{i_j}$. This is because
the equalities added to $\mathcal{L}$ during the first $j$ steps did
not cover $b_{i_{j+1}}$. See Figure~\ref{fig:avoid}. Since
$y_{i_{j+2}}>y_{i_{j+1}}$, we must also have $x_{i_{j+2}} > y_{i_j}$:
otherwise, $\hat{P}_{i_{j+1}}$ would not be the ``rightmost''
constraint for $b_{i_{j+1}}$.  Therefore, the variables in
$I_{i_{j+2}}$ and $I_{i_j}$ do not overlap, and hence no $b_i$ can
appear in more than two equalities in $\mathcal{L}$.
Hence, adding up all of the equalities in $\mathcal{L}$ (and
noting that the $b_i$ are non-negative) we obtain
\[
2\sum_{i=1, \dots, k}b_i \ge
\sum_{j~:~i_j\in\mathcal{L}}c(\hat{P_j}\setminus P).
\]
On the other hand, each equality $I_i$ has a corresponding
inequality based on constraint (2) applied to $\mathbf{b}'$,
namely $b'_{x_{i}+1}+\dots+b'_{y_i} \leq c(\hat{P}_i\setminus P)$.
Summing these inequalities we have $\sum_{i=1, \dots, k}b'_i \le
\sum_{j~:~i_j\in\mathcal{L}}c(\hat{P_j}\setminus P)$. The result
follows from this and the previous expression.
\end{proof}

Finally, we show that for matroids all four payment bounds coincide.

\begin{theorem}\label{thm:uppermatroid}
For any matroid $M=(\mathcal{E}, \mathcal{F})$ with cost vector $\mathbf{c}$,
${\TUMax} / \TUMin = 1$.
\end{theorem}

\begin{proof}
Let $S$, $|S|=k$ be
the le\-xi\-co\-gra\-phi\-cally-least cheapest base of $M$.
We can assume without loss of generality
that $S = \{e_1, \dots, e_k\}$.
Let $\mathbf{b}=(b_1,\dots,b_k)$ and
$\mathbf{b}'=(b'_1,\dots,b'_k)$ be bid vectors that correspond to
$\TUMin$ and $\TUMax$, respectively.
For the bid vector $\mathbf{b}$ and
any $e_i\in S$, consider a constraint in~(2) that is tight for $e_i$
and the base $S'$ that is associated with this constraint.
Suppose $S\setminus S'=\{e_{i_1}, \dots, e_{i_t}\}$,
i.e., the tight constraint for $e_i$ is of the form
 $b_{i_1}+\dots +b_{i_t}=c(S'\setminus S)$, $i\in\{i_{i_1}, \dots, i_{i_t}\}$.
By Proposition~\ref{matroid} there is a mapping $f$ such that
$S'\setminus S=\{f(e_{i_1}), \dots, f(e_{i_t})\}$ and for $j=1, \dots, t$
the set $S\setminus\{e_{i_j}\}\cup\{f(e_{i_j})\}$ is a base.
Therefore by condition~(2) we have $b(e_{i_j})\le c(f(e_{i_j}))$ for all
$j=1, \dots, t$. Consequently, it must be the case  that all these
constraints are tight as well, and in particular we have $b_i=c(f(e_i))$.
On the other hand, as $S\setminus\{e_{i}\}\cup\{f(e_{i})\}\in\mathcal{F}$,
we also have $b'_i\le c(f(e_i))=b_i$. As this holds for any $i=1, \dots, k$,
we have $\TUMax \le \TUMin$. Since also $\TUMin\le \TUMax$, the theorem
follows.
\end{proof}

\section{Truthful mechanisms for vertex cover}\label{sec:vc}

Recall that for a vertex-cover auction on a graph $G=(V, E)$, an
\emph{allocation rule} is an algorithm that takes as input a bid $b_v$
for each vertex $v$ and returns a vertex cover $\hat{S}$ of $G$. As
explained in Section~\ref{sec:preliminaries}, we can combine any
monotone allocation rule with threshold payments to obtain a truthful
auction.

Two natural examples of monotone allocation rules are $VCG$, which
finds an optimal vertex cover, and the mechanism $\mathcal{M}_{GR}$
that uses the greedy allocation algorithm. However, $VCG$ cannot be
guaranteed to run in polynomial time unless $P=NP$ and
$\mathcal{M}_{GR}$ has a worst-case approximation ratio of $\log n$.

Another approximation algorithm for (weighted) vertex cover, which has
approximation ratio 2, is the \emph{local ratio} algorithm
$A_{LR}$~\cite{survey,bye}.  This algorithm considers the edges of $G$
one by one. Given an edge $e=(u,v)$, it computes $\epsilon=\min\{b_u,
b_v\}$ and sets $b_u=b_u-\epsilon$, $b_v=b_v-\epsilon$. After all
edges have been processed, $A_{LR}$ returns the set of vertices
$\{v\mid b_v=0\}$. It is not hard to check that if the order in which
the edges are considered is independent of the bids, then this
algorithm is monotone as well. Hence, we can use it to construct a
truthful auction $\mathcal{M}_{LR}$ that is guaranteed to select a
vertex cover whose cost is within a factor of 2 from the optimal.

However, while the quality of the solution produced by $A_{LR}$ is
much better than that of $\mathcal{M}_{GR}$, we still need to show
that its total payment is not too high. In the next subsection, we
bound the frugality ratio of $\mathcal{M}_{LR}$ (and, more generally,
all algorithms that satisfy the condition of {\em local optimality},
defined later) by $2\Delta$, where $\Delta$ is the maximum degree of
$G$.  We then prove a matching lower bound showing that for some
graphs the frugality ratio of any truthful auction is at least
$\Delta/2$.

\subsection{Upper bound}

For vertices $v$ and $w$, $v\sim w$ means that there is an edge
between $v$ and $w$.
We say that an allocation rule is \emph{locally optimal} if whenever
$b_v > \sum_{w\sim v} b_w$, the vertex~$v$ is not chosen.
Note that for any such rule the threshold bid $t_v$ of~$v$ satisfies
$t_v \leq \sum_{w\sim v} b_w$.

\begin{remark}
The mechanisms $VCG$, $\mathcal{M}_{GR}$, and $\mathcal{M}_{LR}$ are
locally optimal.
\end{remark}

\begin{theorem}\label{thm:2delta}
Any vertex cover auction~$\mathcal{M}$ on a graph with maximum
degree $\Delta$ that has a locally optimal and
monotone allocation rule and pays each agent its threshold bid has
frugality ratio $\phi_{\mathrm{NTUmin}}(\mathcal{M})\leq 2 \Delta$.
\end{theorem}

To prove Theorem~\ref{thm:2delta}, we first show that the total
payment of any locally optimal mechanism does not exceed $\Delta
c(V)$. We then demonstrate that $\NTUMin\ge c(V)/2$. By combining
these two results, the theorem follows.

\begin{lemma}\label{lem:pay}
Let $G=(V, E)$ be a graph with maximum degree $\Delta$.  Let
$\mathcal{M}$ be a vertex-cover auction on $G$ that satisfies the
conditions of Theorem~\ref{thm:2delta}.  Then for any cost vector
$\mathbf{c}$, the total payment of $\mathcal{M}$ satisfies
$p_{\mathcal{M}}({\mathbf c}) \leq \Delta c(V)$.
\end{lemma}

\begin{proof}
First note that any such auction is truthful, so we can assume that
each agent's bid is equal to its cost.  Let $\hat{S}$ be the vertex
cover selected by $\mathcal{M}$.  Then by local optimality
\[
p_{\mathcal{M}}({\mathbf c}) =
\sum_{v\in \hat{S}} t_v \leq
\sum_{v\in\hat{S}}\sum_{w\sim v} b_w =
\sum_{v\in\hat{S}}\sum_{w\sim v} c_w \leq
\sum_{w\in V} \Delta c_w = \Delta c(V).
\]
\end{proof}

We now derive a lower bound on $\TUMax$; while not essential for the
proof of Theorem~\ref{thm:2delta}, it helps us build the intuition
necessary for that proof.

\begin{lemma}\label{lem:VCTUMax}
For a vertex cover instance $G=(V,E)$ in which $S$ is a minimum-cost
vertex cover with respect to cost vector $\mathbf{c}$,
$\TUMaxS\geq c(V\setminus S)$.
\end{lemma}

\begin{proof}
For a vertex~$w$ with at least one neighbour in~$S$, let $d(w)$ denote
the number of neighbours that~$w$ has in~$S$.  Consider the bid vector
$\mathbf{b}$ in which, for each $v\in S$, $b_v = \sum_{w\sim
v,w\not\in S} c_w/d(w)$.  Then $\sum_{v\in S} b_v = \sum_{v\in S}
\sum_{w\sim v,w\not\in S} c_w/d(w) = \sum_{w\notin S} c_w =
c(V\setminus S)$.  To finish we want to show that $\mathbf{b}$ is
feasible in the sense that it satisfies (2).  Consider a vertex cover
$T$, and extend the bid vector $\mathbf{b}$ by assigning $b_v=c_v$ for
$v\notin S$.  Then
\[
b(T) = c(T\setminus S) + b(S\cap T) \geq c(T\setminus S) +
\sum_{v\in S\cap T} \sum_{w\in \overline{S}\cap \overline{T}:w\sim v}
c_w/d(w),
\]
and since all edges between $\overline{S}\cap
\overline{T}$ and~$S$ go to $S\cap T$, the right-hand-side is equal to
\[
c(T\setminus S) + \sum_{w\in \overline{S}\cap \overline{T}} c_w =
c(T\setminus S)+c(\overline{S}\cap\overline{T})=c(V\setminus S) =b(S).
\]
\end{proof}

Next, we prove a lower bound on $\NTUMaxS$; we will then use it to
obtain a lower bound on $\NTUMin$.

\begin{lemma}\label{lem:VCNTUMax}
For a vertex cover instance $G=(V,E)$ in which $S$ is a minimum-cost
vertex cover with respect to cost vector $\mathbf{c}$,
$\NTUMaxS\geq c(V\setminus S)$.
\end{lemma}

\begin{proof}
If $c(S)\ge c(V\setminus S)$, by condition~(1) we are done.
Therefore, for the rest of the proof we assume that $c(S)<c(V\setminus
S)$.  We show how to construct a bid vector $(b_e)_{e\in S}$ that
satisfies conditions~(1) and~(2) such that $b(S)\ge c(V\setminus S)$;
clearly, this implies $\NTUMaxS\ge c(V\setminus S)$.

Recall that a network flow problem is described by a directed graph
$\Gamma=(V_{\Gamma}, E_{\Gamma})$,
a source node $s\in V_\Gamma$, a sink node $t\in V_\Gamma$,
and a vector of capacity constraints $a_e$, $e\in E_\Gamma$.
Consider a network $(V_\Gamma, E_\Gamma)$ such that
$V_\Gamma=V\cup\{s, t\}$,
$E_\Gamma=E_1\cup E_2\cup E_3$, where
$E_1=\{(s, v)\mid v\in S\}$,
$E_2=\{(v, w)\mid v\in S, w\in V\setminus S, (v,w)\in E\}$,
$E_3=\{(w, t)\mid w\in V\setminus S\}$.
Since $S$ is a vertex cover for $G$, no edge of $E$ can have both
of its endpoints in $V\setminus S$, and by construction, $E_2$
contains no edges with both endpoints in $S$. Therefore,
the graph $(V, E_2)$ is bipartite with parts $(S, V\setminus S)$.

Set the capacity constraints for $e\in E_\Gamma$ as follows: $a_{(s,
v)}=c_v$, $a_{(w,t)}=c_w$, $a_{(v, w)}=+\infty$ for all $v\in S$,
$w\in V\setminus S$.  Recall that a {\em cut} is a partition of the
vertices in $V_\Gamma$ into two sets $C_1$ and $C_2$ so that $s\in
C_1$, $t\in C_2$; we denote such a cut by $C=(C_1, C_2)$. Abusing
notation, we write $e=(u,v)\in C$ if $u\in C_1, v\in C_2$ or $u\in
C_2, v\in C_1$, and say that such an edge $e=(u, v)$ {\em crosses} the
cut $C$.  The {\em capacity} of a cut $C$ is computed as ${\rm
cap}(C)=\sum_{(v,w)\in C}a_{(v,w)}$. We have ${\rm cap}(s,
V\cup\{t\})=c(S)$, ${\rm cap}(\{s\}\cup V, t)=c(V\setminus S)$.

Let $C_{\min}=(\{s\}\cup S'\cup W', \{t\}\cup S''\cup W'')$ be a
minimum cut in $\Gamma$, where $S', S''\subseteq S$, $W', W''\subseteq
V\setminus S$. See Figure~\ref{fig:cut}. As ${\rm cap}(C_{\min})\le {\rm cap}(s,
V\cup\{t\})=c(S) <+\infty$, and any edge in $E_2$ has infinite
capacity, no edge $(u, v)\in E_2$ crosses $C_{\min}$.

Consider the network $\Gamma'=(V_{\Gamma'}, E_{\Gamma'})$, where
$V_{\Gamma'}=\{s\}\cup S'\cup W'\cup\{t\}$, $E_{\Gamma'}=\{(u, v)\in
E_\Gamma\mid u,v\in V_{\Gamma'}\}$.  Clearly, $C'=(\{s\}\cup S'\cup
W', \{t\})$ is a minimum cut in $\Gamma'$ (otherwise, there would
exist a smaller cut for $\Gamma$).  As ${\rm cap}(C')=c(W')$, we have
$c(S')\ge c(W')$.

Now, consider the network $\Gamma''=(V_{\Gamma''}, E_{\Gamma''})$,
where $V_{\Gamma''}=\{s\}\cup S''\cup W''\cup\{t\}$,
$E_{\Gamma''}=\{(u, v)\in E_\Gamma\mid u,v\in V_{\Gamma''}\}$.
Similarly, $C''=(\{s\}, S''\cup W''\cup\{t\})$ is a minimum cut in
$\Gamma''$, ${\rm cap}(C'')=c(S'')$. As the size of a maximum flow from $s$
to $t$ is equal to the capacity of a minimum cut separating $s$ and
$t$, there exists a flow ${\cal F}=(f_e)_{e\in E_{\Gamma''}}$ of size
$c(S'')$. This flow has to saturate all edges between $s$ and $S''$,
i.e., $f_{(s,v)}=c_v$ for all $v\in S''$.  Now, increase the
capacities of all edges between $s$ and $S''$ to $+\infty$.  In the
modified network, the capacity of a minimum cut (and hence the size of
a maximum flow) is $c(W'')$, and a maximum flow ${\cal
F}'=(f'_e)_{e\in E_{\Gamma''}}$ can be constructed by greedily
augmenting ${\cal F}$.

Set $b_v=c_v$ for all $v\in S'$, $b_v=f'_{(s,v)}$ for all $v\in S''$.
As ${\cal F}'$ is constructed by augmenting ${\cal F}$, we have
$b_v\ge c_v$ for all $v\in S$, i.e., condition~(1) is satisfied.

\begin{figure}
\begin{center}
\epsfig{file=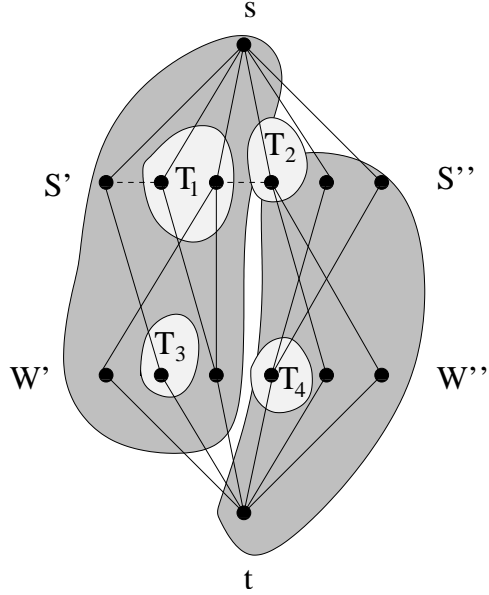, width=2.5in}
\end{center}
\caption{ Proof of Lemma~\ref{lem:VCNTUMax}.
Dashed lines correspond to edges in $E\setminus E_2$}
\label{fig:cut}
\end{figure}

Now, let us check that no vertex cover $T\subseteq V$ can violate
condition~(2).  Set $T_1=T\cap S'$, $T_2=T\cap S''$, $T_3=T\cap W'$,
$T_4=T\cap W''$; our goal is to show that $b(S'\setminus
T_1)+b(S''\setminus T_2) \le c(T_3)+c(T_4)$.  Consider all edges $(u,
v)\in E$ such that $u\in S'\setminus T_1$.  If $(u,v)\in E_2$ then
$v\in T_3$ (no edge in $E_2$ can cross the cut), and if $u,v\in S$
then $v\in T_1\cup T_2$.  Hence, $T_1\cup T_3 \cup S''$ is a vertex
cover for $G$, and therefore $c(T_1)+c(T_3)+c(S'')\ge
c(S)=c(T_1)+c(S'\setminus T_1)+c(S'')$.  Consequently, $c(T_3)\ge
c(S'\setminus T_1)= b(S'\setminus T_1)$.  Now, consider the vertices
in $S''\setminus T_2$.  Any edge in $E_2$ that starts in one of these
vertices has to end in $T_4$ (this edge has to be covered by $T$, and
it cannot go across the cut). Therefore, the total flow out of
$S''\setminus T_2$ is at most the total flow out of $T_4$, i.e.,
$b(S''\setminus T_2)\le c(T_4)$.  Hence, $b(S'\setminus
T_1)+b(S''\setminus T_2) \le c(T_3)+c(T_4)$.
\end{proof}

Finally, we derive a lower bound on the payment bound that is of
interest to us, namely, $\NTUMin$.

\begin{lemma}\label{lem:VCNTUMin}
For a vertex cover instance $G=(V,E)$ in which $S$ is a minimum-cost vertex
cover with respect to cost vector $\mathbf{c}$,
$\NTUMinS\geq c(V\setminus S)$.
\end{lemma}

\begin{proof}
Suppose for contradiction that ${\mathbf c}$ is a cost vector with
minimum-cost vertex cover $S$ and $\NTUMinS<c(V \setminus S)$. Let
${\mathbf b}$ be the corresponding bid vector and let ${\mathbf c}'$
be a new cost vector with $c'_v=b_v$ for $v\in S$ and $c'_v=c_v$ for
$v\not\in S$. Condition~(2) guarantees that $S$ is an optimal solution
to the cost vector ${\mathbf c}'$. Now compute a bid vector ${\mathbf
b}'$ corresponding to $\NTUMaxPrimeS$. We claim that $b'_v=c'_v$ for
any $v\in S$. Indeed, suppose that $b'_v > c'_v$ for some $v\in S$
($b'_v=c'_v$ for $v\not\in S$ by construction).  As ${\mathbf b}$
satisfies conditions (1)--(3), among the inequalities in~(2) there is
one that is tight for $v$ and the bid vector ${\mathbf b}$. That is,
$b(S\setminus T) =c(T\setminus S)$. By the construction of ${\mathbf
c}'$, $c'(S\setminus T)=c'(T\setminus S)$.  Now since $b'_w\geq c'_w$
for all $w\in S$, $b'_v>c'_v$ implies $b'(S\setminus T)>c'(S\setminus
T)=c'(T\setminus S)$.  But this violates (2).  So we now know
${\mathbf b}'={\mathbf c}'$.  Hence, we have $\NTUMaxPrimeS = \sum_{v
\in S} b_v = \NTUMinS<c(V \setminus S)$, giving a contradiction to the
fact that $\NTUMaxPrimeS\geq c'(V\setminus S)$ which we proved in
Lemma~\ref{lem:VCNTUMax}.
\end{proof}

As $\NTUMinS$ satisfies condition~(1), we have $\NTUMinS\geq c(S)$.
Together will Lemma~\ref{lem:VCNTUMin}, this implies $\NTUMinS\geq
\max\{c(V\setminus S),c(S)\}\geq c(V)/2$. Combined with
Lemma~\ref{lem:pay}, this completes the proof of
Theorem~\ref{thm:2delta}.

\begin{remark}
As $\NTUMin\le \NTUMax\le \TUMax$, our bound of $2\Delta$ extends to
the smaller frugality ratios that we consider, i.e.,
$\phi_{\mathrm{NTUmax}}(\mathcal{M})$ and
$\phi_{\mathrm{TUmax}}(\mathcal{M})$.  It is not clear whether it
extends to the larger frugality ratio
$\phi_{\mathrm{TUmin}}(\mathcal{M})$.  However, the frugality ratio
$\phi_{\mathrm{TUmin}}(\mathcal{M})$ is not realistic because the
payment bound $\TUMin$ is inappropriately low---we show in
Section~\ref{sec:properties} that $\TUMin$ can be significantly
smaller than the total cost of a cheapest vertex cover.
\end{remark}

\subsubsection*{Extensions}
We can also apply our results to monotone vertex-cover algorithms that
do not necessarily output locally-optimal solutions.  To do so, we
simply take the vertex cover produced by any such algorithm and
transform it into a locally-optimal one, considering the vertices in
lexicographic order and replacing a vertex $v$ with its neighbours
whenever $b_v > \sum_{u\sim v}b_u$. Note that if a vertex $u$ gets
added to the vertex cover during this process, it means that it has a
neighbour whose bid is higher than $u$'s bid $b_u$, so after one pass all
vertices in the vertex cover satisfy $b_v \le \sum_{u\sim v}b_u$.
This procedure is monotone in bids, and it can only decrease the cost
of the vertex cover. Therefore, using it on top of a monotone
allocation rule with approximation ratio $\alpha$, we obtain a
monotone locally-optimal allocation rule with approximation ratio
$\alpha$.  Combining it with threshold payments, we get an auction
with $\ratioNTUMin\le 2\Delta$.  Since any truthful auction has a
monotone allocation rule, this procedure transforms any truthful
mechanism for the vertex-cover problem into a frugal one while
preserving the approximation ratio.

\subsection{Lower bound}

In this subsection, we prove that the upper bound of
Theorem~\ref{thm:2delta} is essentially optimal. Our proof uses the
techniques of~\cite{ess}, where the authors prove a similar result for
shortest-path auctions.

\begin{theorem}\label{thm:delta/4}
For any $\Delta>0$, there exists a graph $G$ with $2\Delta$ vertices
and degree $\Delta$, such that for any truthful mechanism
$\mathcal{M}$ on $G$ we have $\phi_{\mathrm{NTUmin}}(\mathcal{M})\ge
\Delta/2$.
\end{theorem}

\begin{proof}
Let $G$ be a complete bipartite graph with parts $L$ and $R$,
$|L|=|R|=\Delta$, thus $G$ has degree $\Delta$.

We consider two families of cost vectors for $G$.  Under a cost
vector $\x\in X$, $G$ has one vertex of cost 1; all other vertices
cost 0.  Under a cost vector $\y\in Y$, each of $L$ and $R$ has one
vertex of cost 1, and all other vertices have cost 0.  Clearly,
$|X|=2\Delta$, $|Y|=\Delta^2$.  We construct a bipartite
graph $W$ with the vertex set $X\cup Y$ as follows.

Consider a cost vector $\y\in Y$; let its cost-1 vertices be $v_l\in
L$ and $v_r\in R$.  By changing the cost of either of these vertices
to 0, we obtain a cost vector in $X$. Let $\x_{l}$ and $\x_{r}$ be the
cost vectors obtained by changing the cost of $v_l$ and $v_r$,
respectively.  The vertex cover chosen by $\mathcal{M}(\y)$ must
either contain all vertices in $L$ or all vertices in $R$. In the
former case, we add to $W$ an edge from $\y$ to $\x_l$ and in the
latter case we add to $W$ an edge from $\y$ to $\x_r$ (if the vertex
cover includes all of $G$, $W$ contains both of these edges).

The graph $W$ has at least $\Delta^2$ edges, so there must exist an
$\x\in X$ of degree at least $\Delta/2$. Let $\mathbf{y}_1, \dots,
\mathbf{y}_{\Delta/2}$ be the other endpoints of the edges incident to
$\x$, and for each $i=1,\dots,\Delta/2$, let $v_i$ be the vertex of
$G$ whose cost is different under $\x$ and $\y_i$; note that all $v_i$
are distinct.

It is not hard to see that ${\mathrm{NTUmin}}(\x)\le 1$: the cheapest
vertex cover contains the all-0 part of $G$, and we can satisfy
conditions (1)--(3) by allowing one of the vertices in the all-0 part
of each block to bid 1, while all other vertices in the cheapest
set bid 0.

On the other hand, by monotonicity of $\mathcal{M}$ we have $v_i\in
\mathcal{M}(\x)$ for $i=1,\dots,\Delta/2$ ($v_i$ is in the winning set
under $\y_i$, and $\x$ is obtained from $\y_i$ by decreasing the cost
of $v_i$), and moreover, the threshold bid of each $v_i$ is at least
1, so the total payment of $\mathcal{M}$ on $\x$ is at least
$\Delta/2$.  Hence, $\ratioNTUMin(\mathcal{M})\ge
\mathcal{M}(\x)/{\mathrm{NTUmin}}(\x)\ge \Delta/2$.
\end{proof}

\begin{remark}
Theorem~\ref{thm:delta/4} can be extended to apply to graphs with
degree $\Delta$ of unlimited size: a similar argument applies to any
graph made up of multiple copies of the bipartite graph $G$ in the
proof. The resulting lower bound is still $\Delta/2$, i.e.,
it does not depend on the size of the graph.
\end{remark}

\begin{remark}\label{randommechanisms}
The lower bound of Theorem~\ref{thm:delta/4} can be generalised to
randomised mechanisms, where a randomised mechanism is considered to
be truthful if it can be represented as a probability distribution
over truthful mechanisms.  In this case, instead of choosing the
vertex $\x\in X$ with the highest degree, we put both $(\y, \x_l)$ and
$(\y, \x_r)$ into $W$, label each edge with the probability that the
respective part of the block is chosen, and pick $\x\in X$ with the
highest weighted degree.
\end{remark}

\section{Properties of the payment bounds}
\label{sec:properties}

In this section we consider several desirable properties of payment
bounds and evaluate the four payment bounds proposed in this paper
with respect to them. The particular properties that we are interested
in are the relationship with other reasonable bounds, such as the
total cost of the cheapest set $S$ (Section~\ref{sec:comparecost}), or
the total VCG payment (Section~\ref{sec:comparevcg}). We also consider
independence of the choice of $S$ (Section~\ref{sec:choiceS}),
monotonicity (Section~\ref{sec:nonmon}), computational tractability
(Section~\ref{sec:NPH}).

\subsection{Comparison with total cost of winning set}
\label{sec:comparecost}

The basic property of {\em individual rationality} dictates that the
total payment must be at least the total cost of the selected winning
set. In this section we show that amongst the payment bounds we
consider here, $\TUMin$ may be less than the cost of the winning
set $S$. For such set systems, $\TUMin$ may as a result be too low to be
realistic.

Clearly, $\NTUMax$ and $\NTUMin$ are at least the cost of $S$ due to
condition~(1), and so is $\TUMax$, since
$\TUMax\ge\NTUMax$. However, $\TUMin$ fails this test.  The example of
Proposition~\ref{expath} (part $(iii)$) shows that for path auctions,
$\TUMin$ can be smaller than the total cost by a factor of
2. Moreover, there are set systems and cost vectors for which $\TUMin$
is smaller than the cost of the cheapest set $S$ by a factor of
$\Omega(n)$.  Consider, for example, the vertex-cover auction for the
graph of Proposition~\ref{exvcone} with the costs
$c_{X_1}=\dots=c_{X_{n-2}}=c_{X_{n-1}}=1$, $c_{X_0}=0$.  The cost of a
cheapest vertex cover is $n-2$, and the lexicographically first vertex
cover of cost $n-2$ is $\{X_0, X_1, \dots, X_{n-2}\}$.  The
constraints in~(2) are $b_{X_i}+b_{X_0}\le c_{X_{n-1}}=1$.  Clearly,
we can satisfy conditions~(2) and~(3) by setting
$b_{X_1}=\dots=b_{X_{n-2}}=0$, $b_{X_0}=1$, which means that
$\TUMin\le 1$. This example suggests that the payment bound $\TUMin$
is sometimes too strong to be realistic, since it can be substantially
lower than the cost of a cheapest feasible set.

Note, however,
that this is not an issue for matroid auctions becuase,
for matroids, all four payment bounds have the same value.
The paper~\cite{kkt}
shows that if the feasible sets are the bases of a monopoly-free
matroid, then $\ratioNTUMin(\VCG)=1$. It is not difficult to see
that this is also the case for other payment bounds.
\begin{claim}
For any monopoly-free matroid, we have
$$
\ratioTUMin(\VCG)=\ratioTUMax(\VCG)=\ratioNTUMax(\VCG)=\ratioNTUMin(\VCG)=1.
$$
\end{claim}
\begin{proof}
The claim follows immediately from Theorem~\ref{thm:uppermatroid}.
Alternatively, it is not hard to check that the argument used in~\cite{kkt}
for $\NTUMin$ does not use condition~(1) at all and hence it
works for $\TUMin$ as well.
\end{proof}

\subsection{Comparison with VCG payments}
\label{sec:comparevcg}

Another measure of suitability for payment bounds is that they should
not result in frugality ratios that are less then 1 for well-known
truthful mechanisms\footnote{It is too much to ask that they do not
result in frugality ratios that are less than 1 {\em for all}
mechanisms, since this can typically be subverted by artificial
mechanisms.}.  If this is indeed the case, the payment bound
may be too weak, as it becomes too easy to design mechanisms that
perform well with respect to it. It particular, a reasonable
requirement is that a payment bound should not exceed the total
payment of the classical VCG mechanism.

The following proposition shows that $\NTUMax$, and therefore also
$\NTUMin$ and $\TUMin$, do not exceed the VCG payment
$p_{\VCG}({\mathbf c})$.  The proof essentially follows the argument
of Proposition 7 of~\cite{kkt}.

\begin{proposition}\label{VCGge1}
For any set-system auction, $\ratioNTUMax(\VCG) \ge 1$.
\end{proposition}

\begin{proof}
Let $S$ be a winning set chosen by VCG (hence, a cheapest set).
Suppose $e\in S$. The VCG payment $p_e$ is $\min_{T:e\notin T}
\{c_e+c(T)-c(S)\}$.  Let $T_e$ be a feasible set $T$ which achieves
the minimum, so $p_e = c(T_e)-c(S\setminus\{e\})$. But constraint (2)
gives $b(S\setminus T)\leq c(T\setminus S)$ for all~$T$, so since
$e\notin T_e$, $b_e + b((S\setminus T_e)\setminus \{e\}) \leq
c(T_e\setminus S)$, so
\begin{equation*}
b_e \leq c(T_e\setminus S)-b((S\setminus T_e)\setminus \{e\}).
\tag{$4$}
\end{equation*}
Now by constraint (1), $b((S\setminus T_e)\setminus \{e\})\geq
c((S\setminus T_e)\setminus \{e\})$, so (4) gives
\[
\begin{tabular}{rl}
$b_e$ & $\leq c(T_e\setminus S) - c((S\setminus T_e)\setminus\{e\})$ \\
      & $= c(T_e\setminus S)+c(T_e\cap S)-c((S\setminus T_e)\setminus\{e\})
        - c(T_e \cap S)$ \\
      & $\leq c(T_e) - c(S\setminus\{e\})$ \\
      & $= p_e.$
\end{tabular}
\]
Thus, every winner's payment is at least his bid, so the result
follows.
\end{proof}

Proposition~\ref{VCGge1} shows that none of the payment bounds
$\TUMin$, $\NTUMin$ and $\NTUMax$ exceeds the payment of VCG.
However, the payment bound $\TUMax$ can be larger that the total VCG
payment.  In particular, for the instance in
Proposition~\ref{exvcone}, the VCG payment is smaller than $\TUMax$ by
a factor of $n-2$.  We have already seen that $\TUMax \geq n-2$.  On
the other hand, under VCG, the threshold bid of any $X_i$, $i=1,
\dots, n-2$, is 0: if any such vertex bids above 0, it is deleted from
the winning set together with $X_0$ and replaced with $X_{n-1}$.
Similarly, the threshold bid of $X_0$ is 1, because if $X_0$ bids
above 1, it can be replaced with $X_{n-1}$.  So the VCG payment
is~$1$.

This result is not surprising: the definition of $\TUMax$ implicitly
assumes there is co-operation between the agents, while the
computation of VCG payments does not take into account any interaction
between them.  Indeed, co-operation enables the agents to extract
higher payments under VCG. That is, VCG is not group-strategyproof.
This suggests that as a payment bound, $\TUMax$ may be too liberal, at
least in a context where there is little or no co-operation between
agents.  Perhaps $\TUMax$ can be a good benchmark for measuring the
performance of mechanisms designed for agents that can form coalitions
or make side payments to each other, in particular,
group-strategyproof mechanisms.

Another setting in which bounding $\ratioTUMax$ is still of some
interest is when, for the underlying problem, the optimal allocation
and VCG payments are NP-hard to compute.  In this case, finding a {\em
polynomial-time computable} mechanism with good frugality ratio with
respect to $\TUMax$ is a non-trivial task, while bounding the
frugality ratio with respect to more challenging payment bounds could
be too difficult. To illustrate this point, compare the proofs of
Lemma~\ref{lem:VCTUMax} and Lemma~\ref{lem:VCNTUMax}: both require
some effort, but the latter is much more difficult than the former.

\subsection{The choice of $S$}
\label{sec:choiceS}

All payment bounds defined in this paper correspond to the total bid
of all elements in a cheapest feasible set, where ties are broken
lexicographically.  While this definition ensures that our payment
bounds are well-defined, the particular choice of the draw-resolution
rule appears arbitrary, and one might ask whether our payment bounds are
sufficiently robust to be independent of this choice.  It turns out
that is indeed the case for $\NTUMin$ and $\NTUMax$.
\begin{proposition}
The values of $NTUMinS$ and $\NTUMaxS$ do not depend on the choice of $S$.
\end{proposition}
\begin{proof}
Consider two feasible sets $S_1$ and $S_2$ that have the same cost.  In
the computation of ${{\mathrm{NTUmin}}({\mathbf c},S_1)}$, all
vertices in $S_1\setminus S_2$ would have to bid their true cost,
since otherwise $S_2$ would become cheaper than $S_1$. Hence, any bid
vector for $S_1$ can only have $b_e\neq c_e$ for $e\in S_1\cap S_2$,
and hence constitutes a valid bid vector for $S_2$ (in the context
of ${{\mathrm{NTUmin}}({\mathbf c},S_2)}$) and vice versa.  A
similar argument applies to $\NTUMax$.
\end{proof}

However, for $\TUMin$ and $\TUMax$ this is not the case. For example,
consider the set system
\[
\mathcal{E}=\{e_1, e_2, e_3, e_4, e_5\},
\mathcal{F}=\left\{S_1=\{e_1, e_2\},
                   S_2=\{e_2, e_3, e_4\},
                   S_3=\{e_4, e_5\}\right\}
\]
with the costs $c_1=2$, $c_2=c_3=c_4=1$, $c_5=3$.  The cheapest
sets are $S_1$ and $S_2$.  Now ${{\mathrm{TUmax}}({\mathbf c},S_1)}
\le 4$, as the total bid of the elements in $S_1$ cannot exceed the
total cost of $S_3$. On the other hand, ${{\mathrm{TUmax}}({\mathbf
c},S_2)} \ge 5$, as we can set $b_2=3, b_3=0, b_4=2$.  Similarly,
${{\mathrm{TUmin}}({\mathbf c},S_1)} = 4$, because the equalities
in~(3) are $b_1=2$ and $b_1+b_2=4$. But ${{\mathrm{TUmin}}({\mathbf
c},S_2)} \le 3$, since we can set $b_2=1$, $b_3=2$, $b_4=0$.

\subsection{Negative results for $\NTUMin$ and $\TUMin$}

The results in~\cite{kkt} and our vertex cover results are proved for
the frugality ratio $\ratioNTUMin$.  Indeed, it can be argued that
$\ratioNTUMin$ is the ``best'' definition of frugality ratio, because
among those payment bounds that are at least as large as the cost of
a cheapest feasible set, it is most demanding of the algorithm.
However, $\NTUMin$ is not always the easiest or the most natural
payment bound to work with. In this subsection, we discuss several
disadvantages of $\NTUMin$ (and also $\TUMin$) as compared with
$\NTUMax$ and $\TUMax$.

\subsubsection{Nonmonotonicity}
\label{sec:nonmon}
\footnote{
Simultaneously and independently of our work, Chen and Karlin~\cite{kar}
studied the issue of nonmonotonicity of $\NTUMin$ in much more detail.
While some of our results on $\NTUMin$ are subsumed by their work, we present
our results here as we feel that they are relevant in the context of this paper,
and furthermore, they also apply to $\TUMin$.
}

The first problem with $\NTUMin$ is that it is not monotone with
respect to $\mathcal{F}$, in that it may increase when one adds a
feasible set to $\mathcal{F}$.  (It is, however, monotone in the sense
that a losing agent cannot become a winner by raising its cost.)
Intuitively, a good payment bound should satisfy this monotonicity
requirement, as adding a feasible set increases the competition, so it
should drive the prices down.
Note that this is indeed the case for
$\NTUMax$ and $\TUMax$ since a new feasible set adds a constraint
in~(2), thus limiting the solution space for the respective linear
program (recall Remark~\ref{maxeasy}).

\begin{proposition}\label{clm:nm}
Adding a feasible set to $\mathcal{F}$ can increase $\NTUMin$
and $\TUMin$ by a
factor of $\Omega(n)$, where $n$ is the number of agents.
\end{proposition}

\begin{proof}
Let $\mathcal{E}=\{x,x',y_1,\ldots,y_n,z_1,\ldots,z_n\}$.  Let $Y =
\{y_1,\ldots,y_n\}$, $S = Y \cup \{x\}$, $T_i = Y \setminus \{y_i\}
\cup \{z_i\}$, $i=1, \dots, n$, and suppose that $\mathcal{F}=\{S,
T_1, \ldots, T_n\}$.  The costs are $c_x=0$, $c_{x'}=0$, $c_{y_i}=0$,
$c_{z_i}=1$ for $i=1, \dots, n$.  Note that $S$ is a cheapest
feasible set. For $\mathcal{F}$, the bid vector
$b_{y_1}=\dots=b_{y_n}=0$, $b_x=1$ satisfies (1), (2), and (3), so
$\TUMin\le \NTUMin\leq1$.

Let $S' = Y \cup \{x'\}$.  For $\mathcal{F}\cup\{S'\}$, $S$ is still
the lexicographically-least cheapest set.  Any optimal solution has
$b_x=0$ (by constraint in (2) with $S'$).  Condition (3) for $y_i$
implies $b_x + b_{y_i} = c_{z_i}=1$, so $b_{y_i}=1$ and
$\NTUMin=n$. As all constraints in~(1) are of the form $b_e\ge 0$,
we also have $\TUMin=n$.
\end{proof}

For path auctions, it has been shown~\cite{kar} that $\NTUMin$ is
non-monotone in a slightly different sense, i.e., with respect to
adding a new edge (agent) rather than a new feasible set (a team of
existing agents).  We present that example here for completeness.

\begin{figure}
\begin{center}
\epsfig{file=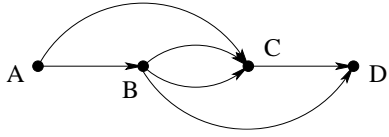, width=2in}
\end{center}
\caption{Nonmonotonicity of $\NTUMin$ for path auctions}
\label{fig:doublediamond}
\end{figure}

\begin{proposition}\label{dd}
For shortest path auctions, adding an edge to the graph can increase
$\NTUMin$ by a factor of $2$.
\end{proposition}

\begin{proof}
Consider the graph of Figure~\ref{fig:diamond} with the edge costs
$c_{AB}=c_{BC}=c_{CD}=0$, $c_{AC}=c_{BD}=1$.  In this graph, $ABCD$ is
the cheapest path, and it is easy to see that $\NTUMin=1$ with the bid
vector $b_{AB}=b_{CD}=0$, $b_{BC}=1$.  Now suppose that we add a new
edge $\widehat{BC}$ of cost 0 between $B$ and $C$, obtaining the graph
of Figure~\ref{fig:doublediamond}.  We can assume that the original
shortest path $ABCD$ is the lexicographically first shortest path in
the new graph, so it gets selected. However, now we have a new
constraint in~(2), namely, $b_{BC}\le c_{\widehat{BC}}=0$, so we have
$\NTUMin=2$ with the bid vector $b_{AB}=b_{CD}=1$, $b_{BC}=0$.
\end{proof}

\begin{remark}
It is not hard to modify the example of Proposition~\ref{dd} so that
the underlying graph has no multiple edges. Also, as all constraints
in~(1) are of the form $b_e\ge 0$, it also applies to $\TUMin$.
\end{remark}

\begin{remark}
We can also show that $\NTUMin$ and $\TUMin$ are
non-monotone for vertex cover.
In this case, adding a new feasible set corresponds to {\em deleting}
edges from the graph. It turns out that deleting a single edge can
increase $\NTUMin$ and $\TUMin$ by a factor of $n-2$; the construction
is based on the graph and the cost vector used in Proposition~\ref{exvcone}.
\end{remark}

\subsubsection{NP-Hardness}
\label{sec:NPH}

Another problem with $\NTUMinS$ is that it is NP-hard to compute, even
if the number of feasible sets is polynomial in $n$.  Again, this puts
it at a disadvantage compared to $\NTUMaxS$ and $\TUMaxS$ (see
Remark~\ref{maxeasy}).

\begin{theorem}\label{thm:NPhard}
Computing $\NTUMin$ is NP-hard, even when the lexicographically-least
cheapest feasible set $S$ is given in the input.
\end{theorem}

\begin{proof}
We reduce {\sc Exact cover by 3-sets}(X3C) to our problem.  An
instance of X3C is given by a universe $G=\{g_1, \dots, g_{n}\}$ and a
collection of subsets $C_1, \dots, C_m$, $C_i\subset G$, $|C_i|=3$,
where the goal is to decide whether one can cover $G$ by $n/3$ of
these sets.  Observe that if this is indeed the case, then each
element of $G$ is contained in exactly one set of the cover.

\begin{lemma}
Consider a minimisation problem ${\cal P}$ of the following form:\\
Minimise $\sum_{i=1, \dots, n} b_i$ under conditions
\begin{itemize}
\item[(1)]
$b_i \ge 0$ for all $i=1, \dots, n$
\item[(2)]
    $\sum_{i \in S_j} b_i \le a_j$ for $j=1,\ldots,k$;
    subsets $S_j\subseteq \{1,\dots,n\}$
\item[(3)]
    for each $b_j$, one of the constraints in (2)
    involving it is tight.
\end{itemize}
For any such ${\cal P}$, one can construct in polynomial time a set
system and a vector of costs $\mathbf{c}$ such that $\NTUMin$ is the
optimal solution to ${\cal P}$.
\label{lem:reword}
\end{lemma}

\begin{proof}
The construction is straightforward: there is an element $e_i$ of cost
0 for each $b_i$, an element $e'_j$ of cost $a_j$ for each $a_j$, the
feasible solutions are $\{e_1, \dots, e_n\}$, or any set obtained from
$\{e_1, \dots,e_n\}$ by replacing the elements indexed by $S_j$,
with $e'_j$.
\end{proof}

By this lemma, all we have to do to prove Theorem~\ref{thm:NPhard} is
to show how to solve X3C by using the solution to a minimisation
problem of the form given in Lemma~\ref{lem:reword}. We do this as
follows.  For each $C_i$, we introduce 4 variables $x_i$, $\bar{x}_i$,
$a_i$, and $b_i$. Also, for each element $g_j$ of $G$ there is a
variable $d_j$.  We use the following set of constraints:

\begin{itemize}
\item
In (1), we have constraints
$x_i\ge 0$, $\bar{x}_i\ge 0$, $a_i\ge 0$, $b_i\ge 0$, $d_j \ge 0$
for all $i=1, \dots, m$, $j=1, \dots, n$.
\item
In (2),
for all $i=1, \dots, m$, we have the following 5 constraints:\\
$x_i+\bar{x}_i \le 1$\\
$x_i      +a_i \le 1$\\
$\bar{x}_i+a_i \le 1$\\
$x_i      +b_i \le 1$\\
$\bar{x}_i+b_i \le 1$.

Also, for all $j=1, \dots, n$ we have the constraint
$d_j + \sum_{i~:~g_j\in C_i}x_i \le 1$.
\end{itemize}
The goal is to minimize $Z = \sum_i(x_i+\bar{x}_i+a_i+b_i)+\sum_j d_j$.

Observe that for each $j$, there is only one constraint involving
$d_j$, so by condition~(3) it must be tight.

Consider the two constraints involving $a_i$. One of them must be
tight; either $x_i+a_i=1$ or $\bar{x}_i+a_i=1$. It follows that
$x_1+\bar{x}_i+a_i+b_i\geq 1$.  Hence, for any feasible solution to
(1)--(3) we have $Z \ge m$.  Now, suppose that there is an exact set
cover. Set $d_j=0$ for $j=1, \dots, n$. Also, if $C_i$ is included in
this cover, set $x_i=1$, $\bar{x}_i=a_i=b_i=0$, otherwise set
$\bar{x}_i=1$, $x_i=a_i=b_i=0$.  Clearly, all inequalities in (2) are
satisfied (we use the fact that each element is covered exactly once),
and for each variable, one of the constraints involving it is
tight. This assignment results in $Z=m$.

Conversely, suppose there is a feasible solution with $Z=m$. As each
addend of the form $x_i+\bar{x}_i+a_i+b_i$ contributes at least 1, we
have $x_i+\bar{x}_i+a_i+b_i=1$ for all $i$, $d_j=0$ for all $j$.  We
will now show that for each $i$, either $x_i=1$ and $\bar{x}_i=0$, or
$x_i=0$ and $\bar{x}_i=1$. For the sake of contradiction, suppose that
$x_i=\delta <1$, $\bar{x}_i=\delta'<1$. As one of the constraints
involving $a_i$ must be tight, we have $a_i\ge \min\{1-\delta,
1-\delta'\}$. Similarly, $b_i\ge \min\{1-\delta, 1-\delta'\}$.  Hence,
$x_i+\bar{x}_i+a_i+b_i \geq \delta+\delta'+2\min\{1-\delta,
1-\delta'\} > 1$. This contradicts the previously noted equality
$x_i+\bar{x}_i+a_i+b_i=1$.

To finish the proof, note that for each $j=1,\dots,n$ we have
$x_{i_1}+\dots + x_{i_k}+d_j = 1$ and $d_j=0$, so the subsets that
correspond to $x_i=1$ constitute a set cover.
\end{proof}

\begin{remark}
In the proofs of Theorem~\ref{thm:NPhard}
all constraints in~(1) are of the form $b_e\ge 0$. Hence, the same
result is true for $\TUMin$.
\end{remark}

\begin{remark}
For shortest-path auctions, the size of $\mathcal{F}$ can be
superpolynomial.  However, there is a polynomial-time separation
oracle for constraints in~(2) (to construct one, use any algorithm for
finding shortest paths), so one can compute $\NTUMax$ and $\TUMax$ in
polynomial time.  On the other hand, recently and independently it was
shown~\cite{kar} that computing $\NTUMin$ for shortest-path auctions
is NP-hard.
\end{remark}

\bigskip
{\bf Acknowledgements.}
We thanks David Kempe for suggesting the ``diamond graph'' auction
and the cost vector used in the proof of Proposition~\ref{expath}[(i)].

\bibliographystyle{abbrv}

\appendix
\section{Nash equilibria and frugality ratios}

Karlin et al.~\cite{kkt}, argue that the payment bound $\nu$ can be
viewed as the total payment in a Nash equilibrium of a certain
game. In this section, we build on this intuition to justify the four
payment bounds introduced above.  We consider two variants of a game
that differ in how profit is shared between the winning players. We
will call these variants the TU game and the NTU game (standing for
``transferable utility'' and ``non-transferable utility''
respectively).  We then show that $\NTUMax$ and $\NTUMin$ correspond
to the worst and the best Nash equilibrium of the NTU game, and
$\TUMax$ and $\TUMin$ correspond to the worst and the best Nash
equilibrium of the TU game. $\NTUMin$ corresponds to the payment bound
$\nu$ of~\cite{kkt}.

In both versions, the players are the elements of the ground set
${\cal E}$.  Each player has an associated cost that is known to all
parties.  The game starts by the buyer selecting a cheapest feasible
set $S\in{\cal F}$ (with respect to the true costs), resolving ties
lexicographically.  Then the elements of $S$ are allowed to make bids,
and the buyer decides whether or not to accept them.  Intuitively,
$S$ ought to be able to win the auction, and we seek bids from $S$
that are low enough to win, and high enough that no member of $S$ has
an incentive to raise his bid (because that would cause him to lose).

Given that $S$ is supposed to win, we modify the game to rule out
behaviour such as elements of $S$ bidding unnecessarily high and
losing. One way to enforce the requirement that $S$ wins is via
fines. If $S$ is not among the cheapest sets with respect to the
bids (where the new cost of a set $T$ is the sum of the total cost of
$T\setminus S$ and the total bid of $T\cap S$), the buyer
rejects the solution and every element of $S$ who bids above its true
cost pays a fine of size $C=\max_{e\in\mathcal{E}}c_e$, while other
elements pay 0. Otherwise, members of $S$ are paid their bids (which
may then be shared amongst members of $S$). This ensures that in a
Nash equilibrium, the resulting bids are never rejected as a result of
$S$ not being the cheapest feasible set.

In the NTU game, we assume that players cannot make payments to each
other, i.e., the utility of each player in $S$ is exactly the
difference between his bid and his true cost. In particular, this
means that no agent will bid below his true cost, which is captured by
condition~(1). In a Nash equilibrium, $S$ is the cheapest set with
respect to the bids, which is captured by condition~(2).  Now, suppose
that condition~(3) is not satisfied for some bidder $e$.  Then the
vector of bids is not a Nash equilibrium: $e$ can benefit from
increasing his bids by a small amount. Conversely, any vector of bids
that satisfies (1), (2) and (3) is a Nash equilibrium: no player wants
to decrease its bid, as it would lower the payment it receives, and no
player can increase its bid, as it would violate~(2) and will cause
this bidder to pay a fine.  As $\NTUMin$ minimises $\sum_{e\in S} b_e$
under conditions (1), (2), and (3), and $\NTUMax$ maximises it, these
are, respectively, the best and the worst Nash equilibrium, from the
buyer's point of view.

In the TU game, the players in $S$ redistribute the profits among
themselves in equal shares, i.e., each player's utility is the
difference between the total payment to $S$ and the total cost of $S$,
divided by the size of $S$. We noted in Section 6.1 that when $S$ is
{\em required} to be the winning set, this may result in Nash
equilibria where members of $S$ make a loss collectively, and not just
individually as a result of condition~(1) not applying.  (Recall that
we do assume that agents' bids are non-negative; condition~$(1^*)$.)
$\TUMin$ thus represents a situation in which ``winners'' are being
coerced into accepting a loss-making contract.

$\TUMax$ does not have the above problem, since it is larger than the
other payment bounds, so members of $S$ will not make a loss.  The
meaning of conditions~(2) and~(3) remains the same: the agents do not
want the buyer to reject their bid, and no agent can improve the
total payoff by raising their bid.  Note that we are not allowing
coalitions (see remark~\ref{no-coalitions}), i.e., coordinated
deviations by two or more players: even though the players share the
profits, they cannot make joint decisions about their strategies.
Similarly to the NTU game, it is easy to see that $\TUMax$ and
$\TUMin$ are, respectively, the worst and the best Nash equilibria of
this game from the buyer's viewpoint.

\begin{remark}\label{no-coalitions}
Allowing payment redistribution within a set is different from
allowing players to form coalitions (as in, e.g., the definition of
strong Nash equilibrium): in the latter case, players are allowed to
make joint decisions about their bids, but they cannot make payments
to each other.
\end{remark}

\end{document}